\documentclass[conference]{IEEEtran}
\IEEEoverridecommandlockouts

\usepackage{textcomp}
\usepackage{xcolor}

\usepackage{amsmath,amssymb,amsfonts}
\usepackage{algorithm}
\usepackage{algorithmic}
\usepackage{graphicx}
\graphicspath{ {Figures/} }
\usepackage{stfloats}
\usepackage{multicol}
\usepackage{multirow}
\usepackage{tablefootnote}
\usepackage[flushleft]{threeparttable}
\usepackage{color, colortbl}
\usepackage{lipsum}

\usepackage{hyperref}
\usepackage{balance}
\usepackage{cite}

\definecolor{Gray}{gray}{0.9}
\definecolor{Highlight}{rgb}{0.99, 0.99, 0.9}

\def\BibTeX{{\rm B\kern-.05em{\sc i\kern-.025em b}\kern-.08em
    T\kern-.1667em\lower.7ex\hbox{E}\kern-.125emX}}
\begin{document}

\title{
pc-COP: An Efficient and Configurable 2048-p-Bit \\ Fully-Connected Probabilistic Computing \\ Accelerator for Combinatorial Optimization
\thanks{\textcopyright $\,$ 2024 IEEE. Personal use of this material is permitted. Permission from IEEE must be obtained for all other uses, in any current or future media, including reprinting/republishing this material for advertising or promotional purposes, creating new collective works, for resale or redistribution to servers or lists, or reuse of any copyrighted component of this work in other works.}
\thanks{A revised version of this paper was published in the proceedings of the 2024 IEEE High Performance Extreme Computing Conference (HPEC) - DOI: \href{https://dx.doi.org/10.1109/HPEC62836.2024.10938509}{10.1109/HPEC62836.2024.10938509}}
}

\author{
\IEEEauthorblockN{Kiran Magar$^{*}$, Shreya Bharathan$^{\dagger}$ and Utsav Banerjee$^{*}$}
\IEEEauthorblockA{
$^{*}$Electronic Systems Engineering, Indian Institute of Science, Bengaluru, India \\
$^{\dagger}$National Institute of Technology, Tiruchirappalli, India \\
Email: kirankailash@iisc.ac.in, shreyabharat2306@gmail.com, utsav@iisc.ac.in
}
}

\maketitle

\begin{abstract}
Probabilistic computing is an emerging quantum-inspired computing paradigm capable of solving combinatorial optimization and various other classes of computationally hard problems.
In this work, we present pc-COP, an efficient and configurable probabilistic computing hardware accelerator with 2048 fully connected probabilistic bits (p-bits) implemented on Xilinx UltraScale+ FPGA.
We propose a pseudo-parallel p-bit update architecture with speculate-and-select logic which improves overall performance by $4 \times$ compared to the traditional sequential p-bit update.
Using our FPGA-based accelerator, we demonstrate the standard G-Set graph maximum cut benchmarks with near-99\% average accuracy.
Compared to state-of-the-art hardware implementations, we achieve similar performance and accuracy with lower FPGA resource utilization.
\end{abstract}


\section{Introduction}
\label{sec:introduction}

Quantum computing \cite{feynman_simulating_1982, nielsen_chuang, shor_quantum_1997, google_supremacy_2019} is a pioneering paradigm in computation which applies the principles of quantum mechanics to revolutionize problem-solving capabilities.
Unique quantum phenomena such as superposition and entanglement enable quantum computers to tackle complex problems with exponential speedup.
While quantum computing holds tremendous potential, it is still in the nascent stages of development and faces significant challenges on its path to practicality and widespread adoption.
A major hurdle is stability and coherence, as quantum systems are highly susceptible to environmental noise and decoherence, which can introduce errors and significantly limit their computational power. Additionally, the development of scalable quantum hardware poses a formidable challenge, requiring advancements in fabrication techniques, error correction methods, cooling systems and the integration of control electronics. Addressing these challenges requires interdisciplinary collaboration and sustained investment in fundamental research and engineering to unlock the transformative capabilities of quantum computing.
These challenges have motivated the emergence of many new physics-inspired computing models such as probabilistic computing \cite{camsari_pbits_2017}, stochastic computing \cite{smithson_stochastic_2019}, simulated annealing \cite{cook_gpu_2019}, probabilistic annealing \cite{jung_annealing_2023} and parallel tempering \cite{zhang_tempering_2024}.
These quantum-inspired algorithms are implemented on classical hardware and draw upon some principles and strategies from quantum computing to significantly enhance the performance of classical algorithms to address specific computational challenges.

\begin{figure}[!t]
\centering
\includegraphics[width=3.4in]{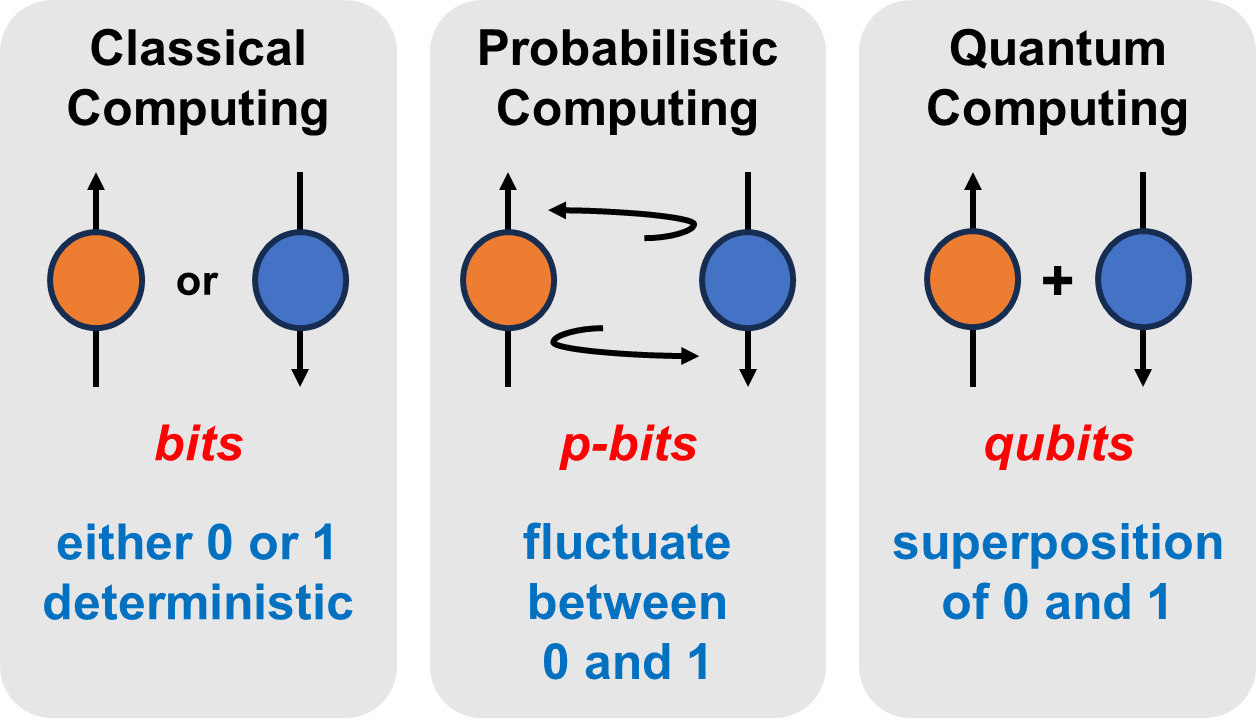}
\caption{Three computing paradigms: classical, probabilistic and quantum \cite{chowdhury_fullstack_2023}.}
\label{fig:computing_paradigms}
\end{figure}

Probabilistic computing is one of the emerging quantum-inspired computing paradigms and it involves the manipulation of unstable stochastic units known as \textit{probabilistic bits} or \textit{p-bits}. Multiple p-bits are interconnected together to construct \textit{probabilistic circuits} or \textit{p-circuits}.
Fig. \ref{fig:computing_paradigms} compares the three computing paradigms - classical (digital), quantum and probabilistic.
The basic building blocks of classical computing are bits which are deterministically either 0 or 1.
The basic building blocks of quantum computing are quantum bits or qubits which can be in a superposition of 0 and 1.
In contrast, probabilistic bits or p-bits rapidly fluctuate between 0 and 1. While qubits require near-absolute-zero temperatures for accurate functionality, such p-bits and p-circuits can be realized at room temperature. Although probabilistic computers are not expected to be a direct replacement for quantum computers, they enable many novel applications at the intersection of classical and quantum computing using existing as well as emerging hardware technologies \cite{camsari_dialogue_2021, chowdhury_fullstack_2023}.
Unlike classical bits which are deterministically either 0 or 1 and quantum bits (qubits) which can exist in a superposition of 0 and 1, p-bits rapidly fluctuate between 0 and 1. While qubits require near-absolute-zero temperatures for accurate functionality, such p-bits and p-circuits can be realized at room temperature, thus enabling many novel applications at the intersection of classical and quantum hardware using existing as well as emerging technologies \cite{camsari_dialogue_2021, chowdhury_fullstack_2023}.
Recent literature has demonstrated the immense potential of probabilistic computing using various software and hardware implementation platforms such as micro-controllers \cite{pervaiz_emulation_2017}, general purpose micro-processors (CPUs) and graphics processing units (GPUs) \cite{khan_maxcut_2022, onizawa_fastconv_2023, onizawa_enhanced_2024}, field-programmable gate arrays (FPGAs) \cite{pervaiz_fpga_2019}, magnetic tunnel junctions (MTJs) \cite{borders_factorization_2019}, resistive random-access memories (RRAMs) \cite{liu_probabilistic_2022}, ferro-electric field-effect transistors (FeFETs) \cite{luo_ferroelectric_2023} and threshold switch devices (TSDs) \cite{heo_oscillation_2023}.
However, these hardware implementations have been limited to either small-scale p-circuits using emerging nano-devices or preliminary architectures using FPGAs. They have limited circuit-level analysis, thus leaving plenty of room for design space exploration and algorithm-specific architectural improvement.

Combinatorial optimization \cite{lucas_ising_2014, mohseni_ising_2022} is an important class of hard problems which can be solved efficiently using probabilistic computing. In this work, we present pc-COP, an efficient and configurable probabilistic computing hardware accelerator with 2048 fully connected p-bits implemented on state-of-the-art Xilinx UltraScale+ FPGA \cite{xilinx_ultrascale}. It is capable of solving large-scale graph maximum cut combinatorial optimization problems \cite{cook_gpu_2019, toshiba_blog_2019} with high accuracy.
Our logarithmic adder tree design for sum-of-products computation boosts overall performance. We approximate the activation function and tune the precision of the annealing schedule to reduce FPGA resource utilization. Our proposed pseudo-parallel p-bit update architecture with speculate-and-select logic improves performance by $4 \times$ compared to the traditional sequential p-bit update. We implement pc-COP on a Xilinx Zynq UltraScale+ MPSoC ZCU104 Evaluation Board and demonstrate near-99\% average accuracy across various G-Set maximum cut benchmarks up to 2,000 nodes \cite{stanford_gset}.

The rest of this paper is organized as follows: Section \ref{sec:background} summarizes the theory of probabilistic computing and its application to solving combinatorial optimization problems such as graph max-cut. Section \ref{sec:architecture} describes the details of our proposed accelerator hardware architecture. Section \ref{sec:implementation} presents detailed FPGA-based implementation results and Section \ref{sec:conclusion} provides concluding remarks and future directions.

\section{Background}
\label{sec:background}

\subsection{Probabilistic Computing}

\begin{algorithm}[!t]
\caption{Overview of p-circuit operation \cite{camsari_pbits_2017, camsari_pbits_2019}}
\label{algo:probabilistic_computing}
\begin{algorithmic}[1]
\REQUIRE number of p-bits $N_m$, interconnection weight matrix \textbf{J} = $[J_{i,j}]_{N_m \times N_m}$ and bias vector \textbf{h} = $[h_i]_{N_m \times 1}$ for application-specific p-circuit, number of samples $N_s$
\ENSURE final p-bit state $m$
\STATE define p-bit state $m = \{m_i \, | \, 1 \le i \le N_m \}$
\STATE randomly initialize p-bits $m_i \in \{ -1, +1 \} \, \forall \, 1 \le i \le N_m$
\FOR{($s = 1$; $s \le N_s$; $s = s + 1$)}
\FOR{($i = 1$; $i \le N_m$; $i = i + 1$)}
\STATE $I_i \leftarrow \beta \times ( \, h_i + \sum_{j=1}^{N_m} \, J_{i,j} \, m_{j} \, )$
\STATE $m_i \leftarrow sgn ( \, rand( -1, +1) + tanh( \, I_i \, ) \, )$
\ENDFOR
\ENDFOR
\RETURN $m = \{m_i \, | \, 1 \le i \le N_m \}$
\end{algorithmic}
\end{algorithm}

\begin{figure}[!t]
\centering
\includegraphics[width=3.4in]{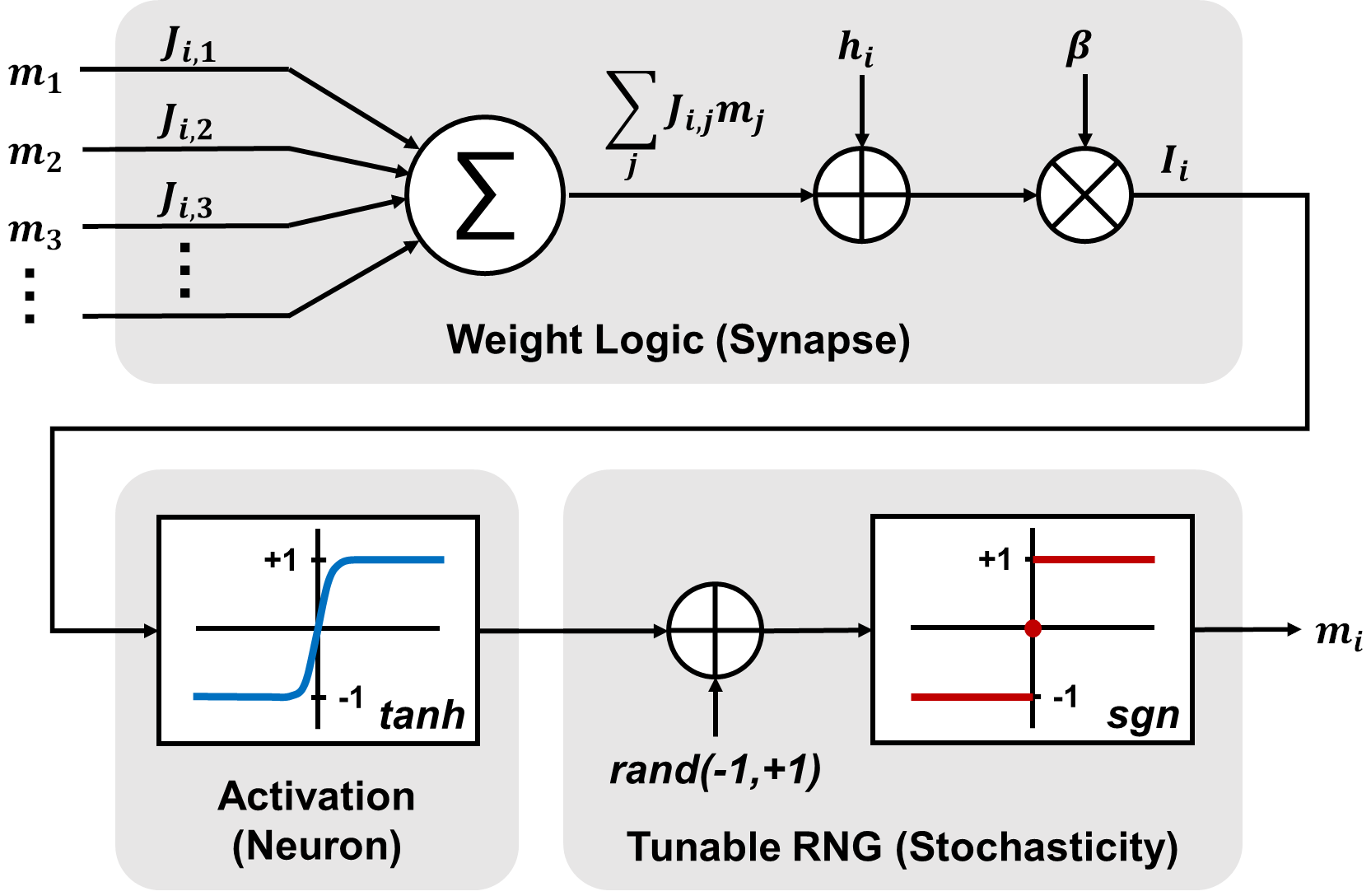}
\caption{Overview of p-bit operation as a binary stochastic neuron \cite{jain_tyche_2023}.}
\label{fig:binary_stochastic_neuron}
\end{figure}

Probabilistic computing is an emerging quantum-inspired computing paradigm capable of solving many interesting problems such as combinatorial optimization, machine learning, quantum emulation and integer factorization \cite{camsari_pbits_2017, camsari_pbits_2019, sutton_autonomous_2020, aadit_computing_2021, kaiser_probabilistic_2021}.
The operation of a p-bit can be represented as a \textit{binary stochastic neuron}, as shown in Fig. \ref{fig:binary_stochastic_neuron}.
The operation of a p-circuit with several p-bits is described in Algorithm \ref{algo:probabilistic_computing}.
The system state is defined as $m = \{m_i \, | \, 1 \le i \le N_m \}$, where each $m_i \in \{ -1, +1 \}$ denotes the corresponding p-bit value.
Each p-bit $m_i$ is updated sequentially based on the weights $J_{i,j}$ and bias values $h_i$. A global constant $\beta$ is used to control the overall strength of p-bit interconnections. A complete sequence of updating all the $N_m$ p-bits in a p-circuit is referred to as a \textit{sample}.
This process is repeated multiple times, thus generating $N_s$ samples.
The stochastic neural network representation is similar to Boltzmann machines \cite{camsari_pbits_2017}. The p-bit update equation ($I_i = \beta \times ( \, h_i + \sum_{j=1}^{N_m} \, J_{i,j} \, m_{j} \, )$) resembles Ising machines \cite{sutton_autonomous_2020} while the sequential nature of the update resembles the iterative evolution in Gibbs sampling \cite{aadit_massively_2022}.
The energy of the system with state $m$ is defined as $E(\{m\}) = - ( \, \sum_{i < j} \, J_{i,j} m_{i} m_{j} + \sum \, h_{i} m_{i} \, )$ which again resembles the quadratic energy model in Ising machines \cite{aadit_massively_2022}. The inherent stochasticity and non-linear activation function in each p-bit, which is a unique feature of probabilistic computing, ensures that various states $m$ are visited according to their corresponding Boltzmann probability $p_{\{m\}} \propto exp[ \, -\beta E(\{m\}) \, ]$, where $\beta$ acts as an \textit{inverse pseudo-temperature} which can enhance or suppress probabilities based on energy minima.
The system of p-bits evolves over consecutive samples to converge towards a low-energy state corresponding to an optimum or near-optimal solution of the problem encoded in the p-circuit. The value of $\beta$ can be tuned across samples to achieve better convergence, which bears resemblance to simulated annealing \cite{sutton_autonomous_2020, aadit_massively_2022}.

Recent literature has explored the use of emerging technologies to efficiently realize the p-bit functionality in hardware \cite{camsari_pbits_2017, camsari_pbits_2019, borders_factorization_2019, liu_probabilistic_2022, grimaldi_annealing_2022, chowdhury_fullstack_2023, heo_oscillation_2023, luo_ferroelectric_2023}. But, large-scale implementations of p-circuits using such emerging nano-devices are yet to be demonstrated experimentally. Therefore, FPGA-based implementations offer a promising near-term alternative. Preliminary FPGA-based architectures have been demonstrated in \cite{pervaiz_fpga_2019, sutton_autonomous_2020, aadit_massively_2022, aadit_tempering_2023, jain_tyche_2023}. However, detailed  circuit-level analysis and architectural optimizations for probabilistic computing are yet to be explored.
In this work, we bridge this gap by providing comprehensive analysis of the compute and memory bottlenecks, resource usage on state-of-the-art FPGA, algorithm-architecture co-optimizations, design space exploration, hardware implementation and experimental results.

\subsection{Combinatorial Optimization Problems}

Combinatorial optimization is a sub-field of mathematical optimization which involves finding the optimal solution out of a finite but large set of possibilities where exhaustive search is intractable \cite{lucas_ising_2014, mohseni_ising_2022}.
Combinatorial optimization is a sub-field of mathematical optimization which involves finding the optimal solution out of a finite set of possibilities by optimizing some objective function. Examples of combinatorial optimization problems (COPs) include the graph maximum cut problem, the traveling salesman problem, the minimum spanning tree problem and the knapsack problem \cite{lucas_ising_2014, mohseni_ising_2022}. Solving COPs is fundamental to various important real-world applications such as VLSI design, machine learning, bioinformatics, telecommunications, software engineering, finance and supply chain management.
Solving large-scale COPs using exhaustive search is intractable, thus necessitating specialized techniques such as dynamic programming, approximation algorithms, metaheuristics, hill climbing and simulated annealing. Ising machines have gained significant interest due to their ability to efficiently compute optimal or near-optimal solutions to such problems \cite{lucas_ising_2014, mohseni_ising_2022}.
In this work, we explore the efficacy of probabilistic computing in solving the prototypical combinatorial optimization problem (COP) of graph maximum cut, also known as \textit{max-cut} \cite{cook_gpu_2019}. The objective of the max-cut problem is to partition the vertices $V$ of a graph $G = (V, E)$ into two complementary sets $S$ and $T$ such that the number of edges ($\in E$) between $S$ and $T$ is as large as possible. The corresponding p-circuit can be constructed by assigning a p-bit $m_i$ corresponding to each vertex $v_i \in V$. Then, the optimal max-cut solution will result in $m_i = +1$ if $v_i \in S$ and $m_i = -1$ if $v_i \in T$ such that the objective function $\sum_{i < j} \, w_{i,j} m_i m_j$ is minimized, where $w_{i,j}$ is the weight of the edge connecting vertices $v_i$ and $v_j$ \cite{toshiba_blog_2019}. Therefore, the p-bit interaction coefficients are obtained as $J_{i,j} = -w_{i,j}$ and $h_i = 0$.
The Stanford G-Set benchmark dataset \cite{stanford_gset}, containing various random, toroidal and planar graphs, is typically used to evaluate max-cut solver implementations. G-Set contains graphs with $w_{i,j} \in \{ 0, 1 \}$ as well as $w_{i,j} \in \{ -1, 0, +1 \}$, therefore 2-bit interaction coefficients $J_{i,j}$ are required.
The Stanford G-Set benchmark dataset \cite{stanford_gset}, containing various random, toroidal and planar graphs with 800, 1,000, 2,000, 3,000, 5,000, 7,000, 8,000, 9,000, 10,000, 14,000 and 20,000 vertices, is typically used to evaluate max-cut solver implementations. G-Set contains graphs with $w_{i,j} \in \{ 0, 1 \}$ as well as $w_{i,j} \in \{ -1, 0, +1 \}$, so we need 2-bit interaction coefficients $J_{i,j}$.%
This work presents an FPGA-based 2048-p-bit probabilistic computing accelerator demonstrating near-99\% average accuracy across G-Set max-cut benchmarks up to 2,000 nodes.

\section{Hardware Architecture}
\label{sec:architecture}

\begin{figure}[!t]
\centering
\includegraphics[width=3.4in]{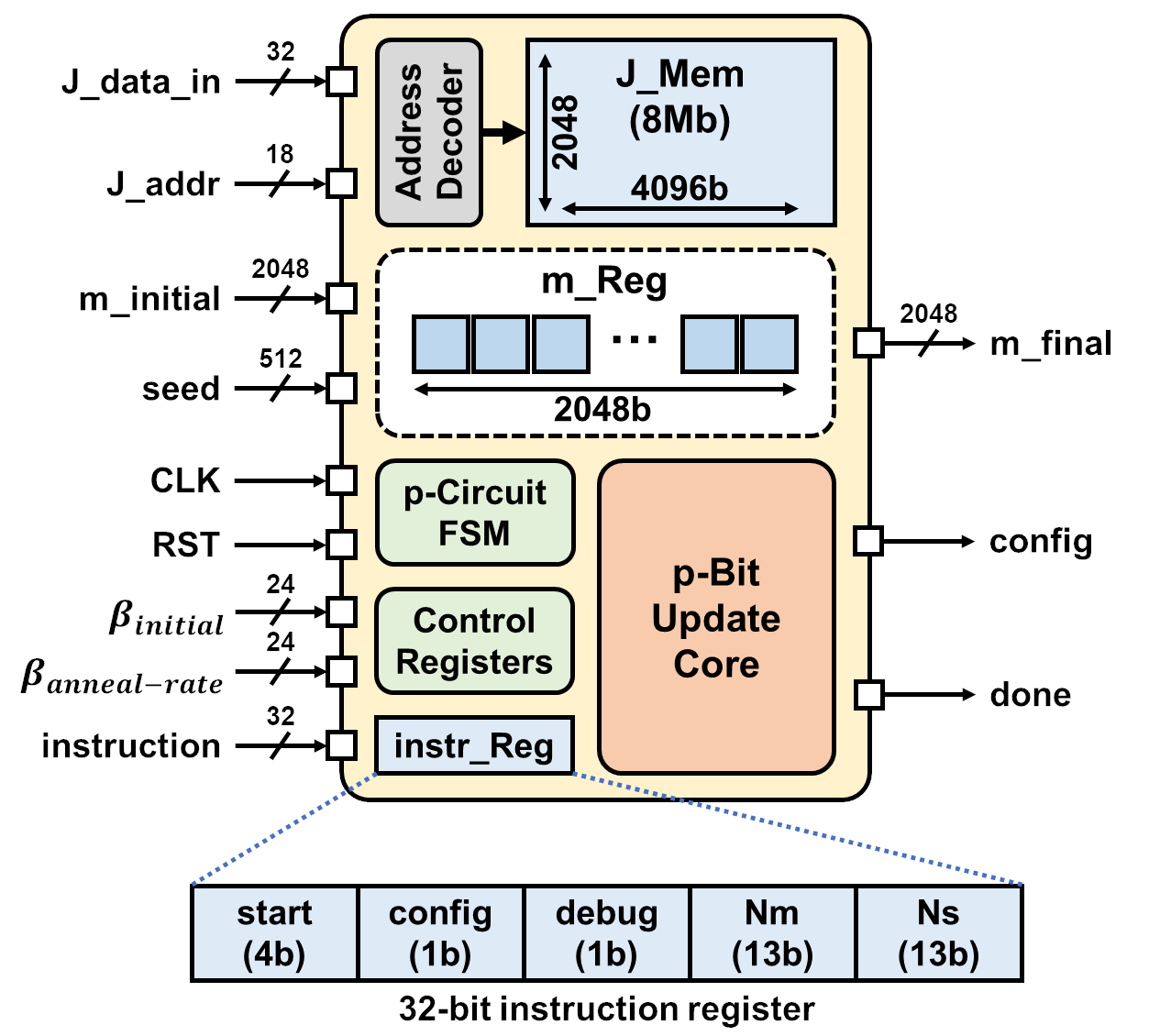}
\caption{Top-level architecture of the proposed pc-COP accelerator.}
\label{fig:top_arch}
\end{figure}

Fig. \ref{fig:top_arch} shows the top-level architecture of pc-COP, our proposed hardware accelerator for solving large-scale max-cut COPs using probabilistic computing. It supports max-cut instances up to 2048 nodes using 2048 p-bits stored in the 2048-bit register m\_Reg, where -1 and +1 p-bit values are encoded as 0 and 1 respectively. According to step 2 of Algorithm \ref{algo:probabilistic_computing}, the initial random state of the p-bit register m\_Reg is configured using a 2048-bit external input. The corresponding $2048 \times 2048$ matrix \textbf{J} of 2-bit interaction coefficients (-1, 0 and + 1 encoded as 11, 00 and 01 respectively) is stored in the $2048 \times 4096$-bit = 8~Mb memory J\_Mem. Although J\_Mem is implemented using 256 physical Block RAM (BRAM) slices of 32~Kb each (excluding error correction coding bits) in FPGA, they are organized such that an entire 4096-bit row of \textbf{J} can be read in a single cycle, as required in step 5 of Algorithm \ref{algo:probabilistic_computing}. The J\_Mem memory is configured with the problem-specific \textbf{J} matrix 32 bits at a time using the 18-bit address input and an address decoder. The functionality of Algorithm \ref{algo:probabilistic_computing} is implemented in the \textit{p-Bit Update Core} and managed by a finite state machine (FSM) and control registers.
A 512-bit seed input is used to configure the stochasticity in the p-bit update circuitry. The 24-bit inputs $\beta_{initial}$ and $\beta_{anneal-rate}$ are used to configure the inverse pseudo-temperature $\beta$.
A 32-bit instruction input is used to program the accelerator as well as configure the number of p-bits $N_m \le 2048$ and the number of samples $N_s$.
The instruction format is shown in Fig. \ref{fig:top_arch} and its length is set to 32 bits to conform with the 32-bit interface between the processing system (PS) and the programmable logic (PL) in Xilinx Zynq MPSoC as discussed in Section \ref{sec:implementation}.
After the accelerator completes $N_s$ samples of the specified max-cut instance with $N_m$ nodes, the final state of the p-bit register is available as output along with several status bits.

\subsection{Inverse Pseudo-Temperature and Annealing Schedule}
\label{subsec:hyperparams}

Algorithm \ref{algo:probabilistic_computing} is controlled by two key hyper-parameters: the inverse pseudo-temperature $\beta$ and the number of samples $N_s$ \cite{sutton_autonomous_2020, aadit_massively_2022}.
Instead of keeping $\beta$ constant, it has been shown that slowly increasing $\beta$ (equivalent to gradual decrease in the pseudo-temperature) significantly improves the accuracy of the system by helping it reach low energy states corresponding to optimal or near-optimal solutions \cite{sutton_autonomous_2020, aadit_massively_2022}.
In our implementation, the inverse pseudo-temperature 
for each sample (each iteration in step 3 of Algorithm \ref{algo:probabilistic_computing}) is obtained according to $\beta_s = \beta_{initial} \times \beta_{anneal-rate}^{s-1} \,\,\, \text{for} \,\,\, 1 \le s \le N_s$ which gives the \textit{annealing schedule}. Based on the number of samples $N_s$, these hyper-parameters are tuned according to the analysis presented in \cite{onizawa_enhanced_2024, onizawa_hyperparameter_2023}. For $N_s = 1000$, we use $\beta_{initial} = 0.01$ and $\beta_{anneal-rate} = 1.005$. For $N_s = 100$, we use $\beta_{initial} = 0.01$ and $\beta_{anneal-rate} = 1.05$.
Our accelerator uses a 24-bit register to store the value of $\beta_s$ with a 4-bit integer part and a 20-bit fractional part.
Based on Python-based software simulation of Algorithm \ref{algo:probabilistic_computing}, we observed that average accuracy remains almost the same for various fractional bit precision of $\beta$ ranging from 4-bit to 32-bit. We chose the fractional bit precision of $\beta$ as 20-bit for our design as it requires the least number of DSP slices (2 DSP multipliers) for implementing the annealing schedule in FPGA.

\subsection{Logarithmic Adder Tree}
\label{subsec:adder_tree}

\begin{figure}[!t]
\centering
\includegraphics[width=3.4in]{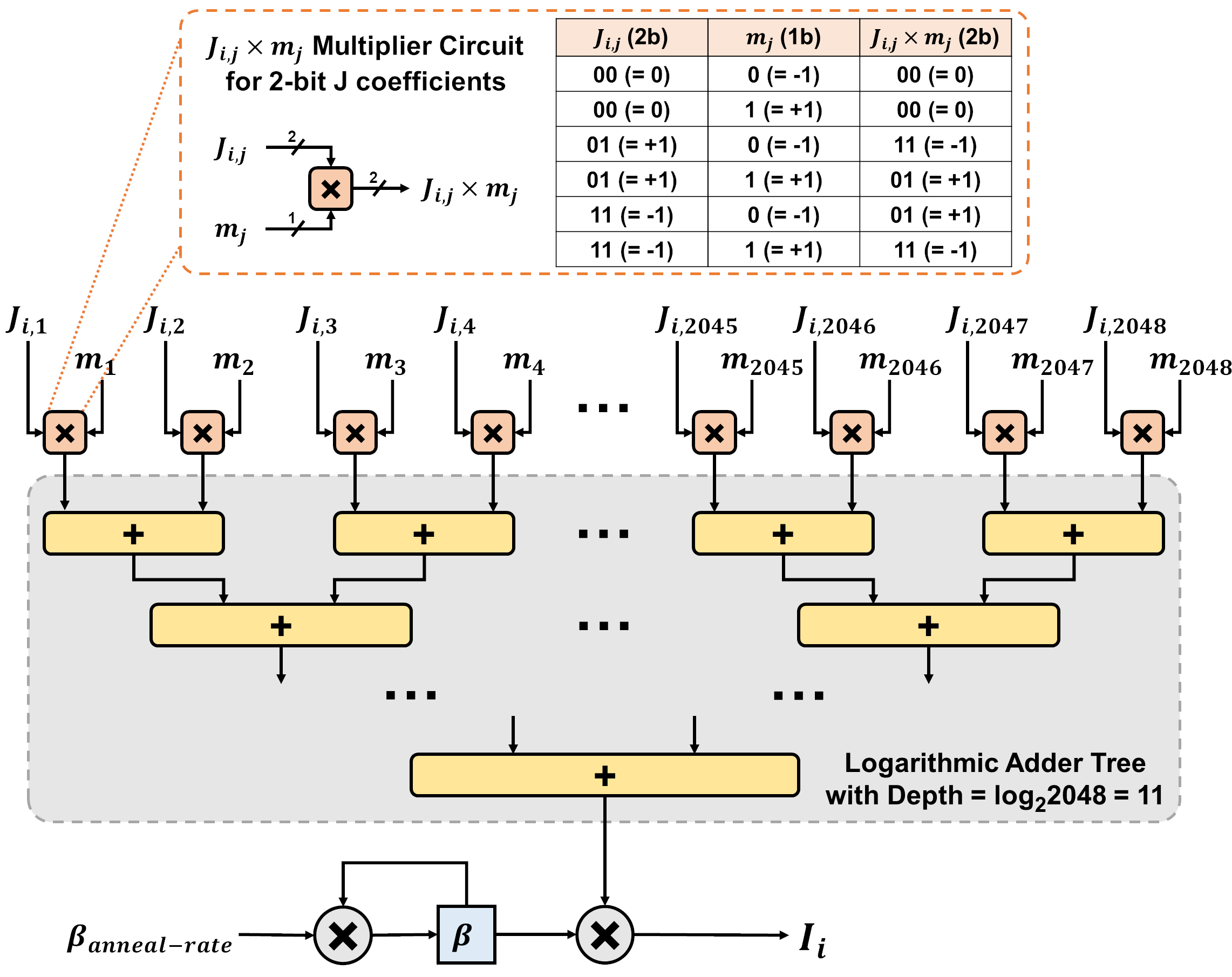}
\caption{Logarithmic adder tree and multiplier circuits for p-bit weight logic.}
\label{fig:log_adder_tree}
\end{figure}

A logarithmic adder tree and associated multiplier circuitry implements step 5 of Algorithm \ref{algo:probabilistic_computing}. For 2-bit interaction coefficients, the $J_{i,j} \times m_{j}$ multiplication is implemented using simple Boolean logic and the corresponding truth table is shown in Fig. \ref{fig:log_adder_tree}.
Using a logarithmic adder tree \cite{jain_tyche_2023} instead of cascaded adders helps reduce the critical path delay by two orders of magnitude \cite{rabaey_chandrakasan, weste_harris}. The output of the adder tree is then multiplied with the inverse pseudo-temperature $\beta$ ($= \beta_s$ for the $s$-th sample) to get $I_i = \beta \times ( \, \sum_{j=1}^{N_m} \, J_{i,j} \, m_{j} \, )$. The annealing schedule from Section \ref{subsec:hyperparams} is implemented using another multiplier which updates the value of $\beta$ after each sample, as shown in Fig. \ref{fig:log_adder_tree}.

\subsection{Activation Function and Random Number Generation}
\label{subsec:activation}

\begin{figure}[!t]
\centering
\includegraphics[width=3.4in]{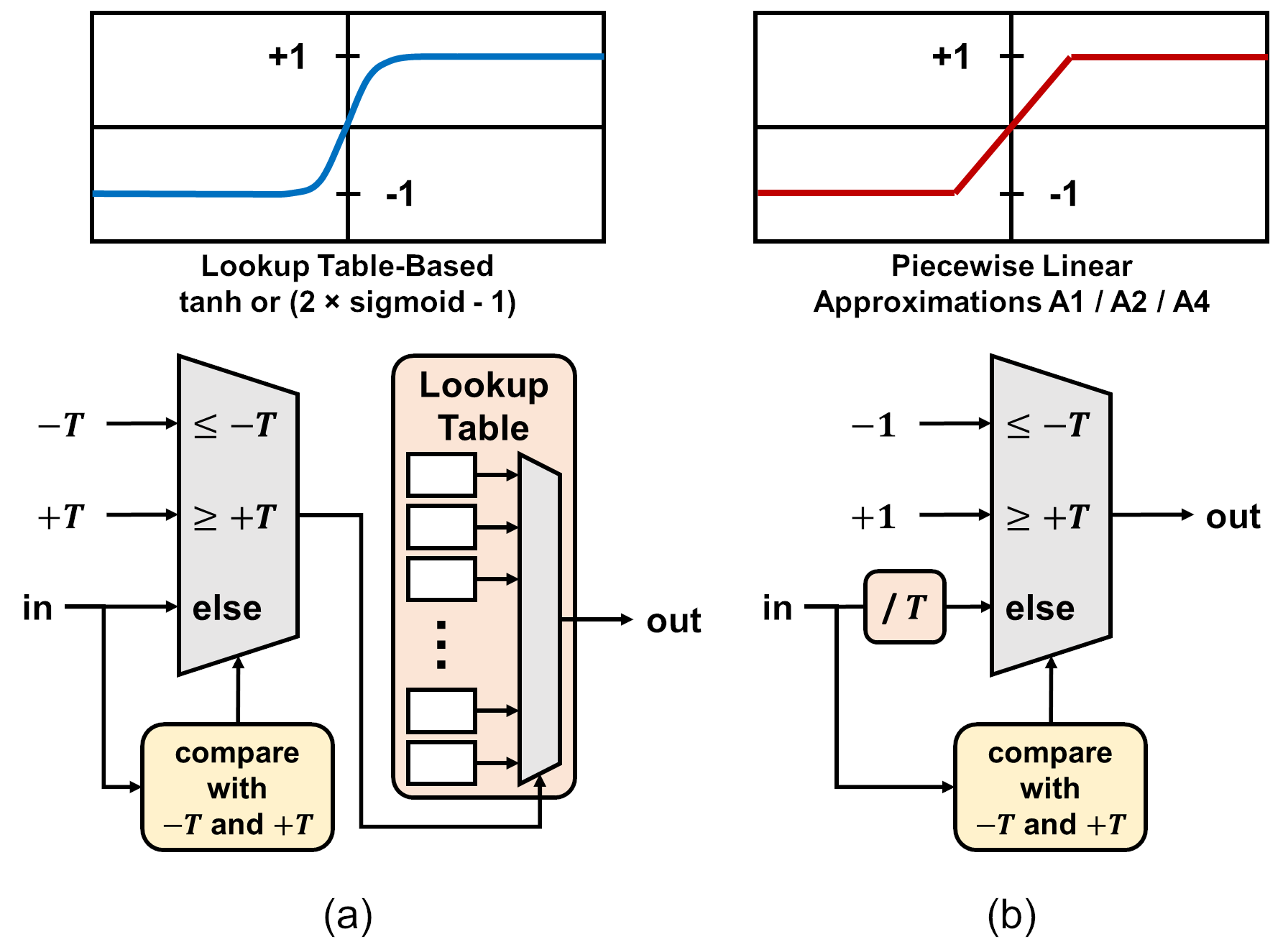}
\caption{Implementations of the activation function: (a) lookup table-based $tanh$ and $2 \times sigmoid - 1$ (threshold $T = 4$), and (b) piece-wise linear approximations $A_1$, $A_2$ and $A_4$ (threshold $T$ = 1, 2 and 4 respectively).}
\label{fig:activation_function_1}
\end{figure}

\begin{figure}[!t]
\centering
\includegraphics[width=3.4in]{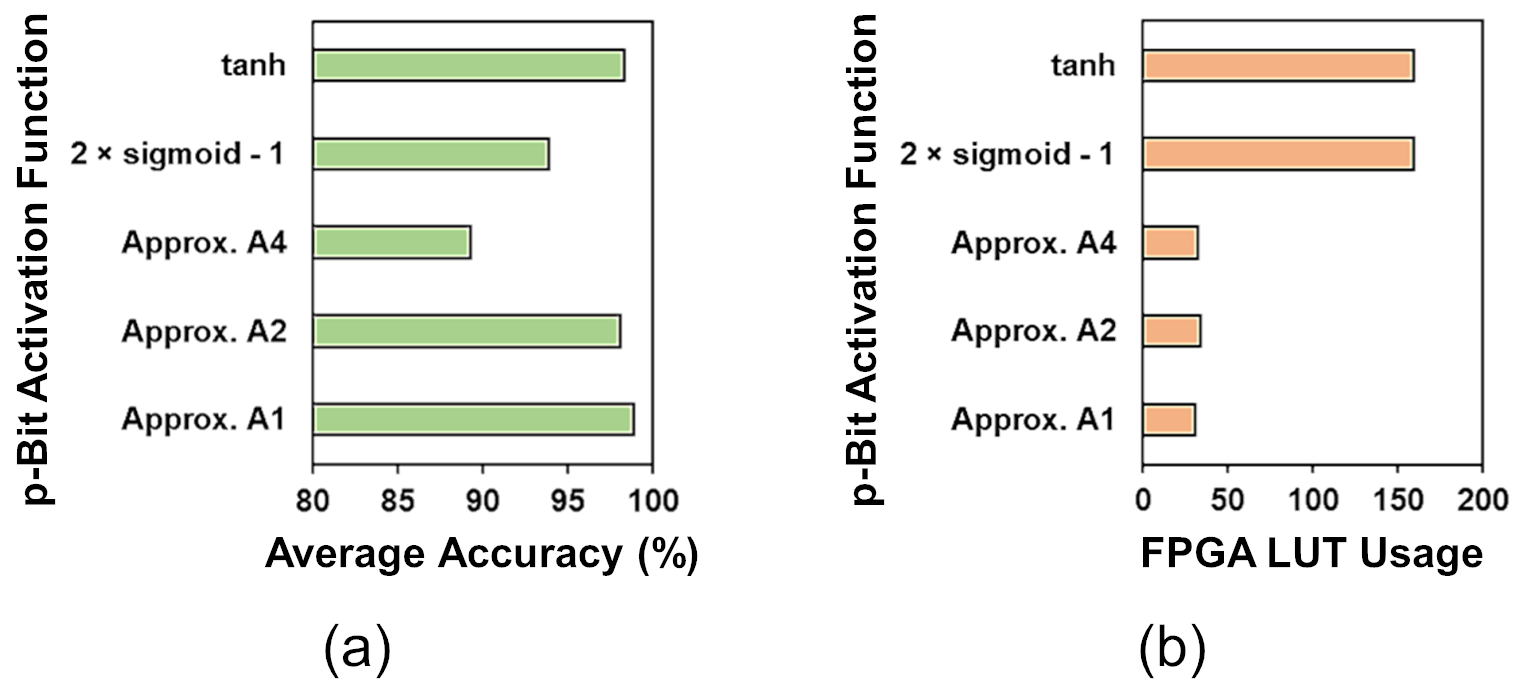}
\caption{Comparison of (a) average accuracy and (b) FPGA resource utilization of various activation function implementations: lookup table-based $tanh$ and $2 \times sigmoid - 1$, and piece-wise linear approximations $A_1$, $A_2$ and $A_4$ (average accuracy obtained from Python-based software simulation with 100 trials of G1, G6, G11, G14 and G18 max-cut benchmarks as examples).}
\label{fig:activation_function_2}
\end{figure}

Previous work \cite{pervaiz_fpga_2019, jain_tyche_2023} has implemented the $tanh$ activation function using a lookup table.
A comparator and a multiplexor are together used to restrict the input range within a certain threshold $[-T, +T]$, where $T = 4$ is typically used for the $tanh$ function.%
In this work, we have explored various approximations of the activation function.
Fig. \ref{fig:activation_function_1} shows the circuit diagrams for lookup table and piece-wise linear approximation implementations. For lookup table, both $tanh$ and its approximation using $2 \times sigmoid - 1$ are analyzed, and the lookup tables consist of 1024 entries with 20-bit fractional precision (consistent with the discussion in Section \ref{subsec:hyperparams}). The piece-wise linear approximations of the activation function are implemented as:
\[
activation \,\, output \approx \begin{cases} 
    -1 & \text{if} \,\, input \le -T \\
    input \, / \, T & \text{if} \,\, -T < input < +T \\
    +1 & \text{if} \,\, input \ge +T
\end{cases}
\]
where division by the threshold $T \in \{1, 2, 4\}$ can be easily implemented using bit-shifts (only wiring).
We observed that the piece-wise linear approximation with $T = 1$ requires the least number of FPGA LUTs ($\approx 5\times$ smaller than lookup table-based implementation) while achieving accuracy similar to the original $tanh$ activation function.
Fig. \ref{fig:activation_function_2} compares the FPGA resource utilization of all these implementations along with their average accuracy (from Python-based software simulation) across various types of max-cut instances. Clearly, approximation $A_1$ requires the least number of FPGA LUTs ($\approx 5\times$ smaller than lookup table-based implementation) while achieving accuracy similar to the original $tanh$ activation function.
Therefore, we implement this simple approximation of the activation function throughout our design. The output size of the activation function, which always lies in $[-1, +1]$, is 22-bit with 1 sign bit, 1 integer bit and 20 fractional bits.

\begin{figure}[!t]
\centering
\includegraphics[width=3.4in]{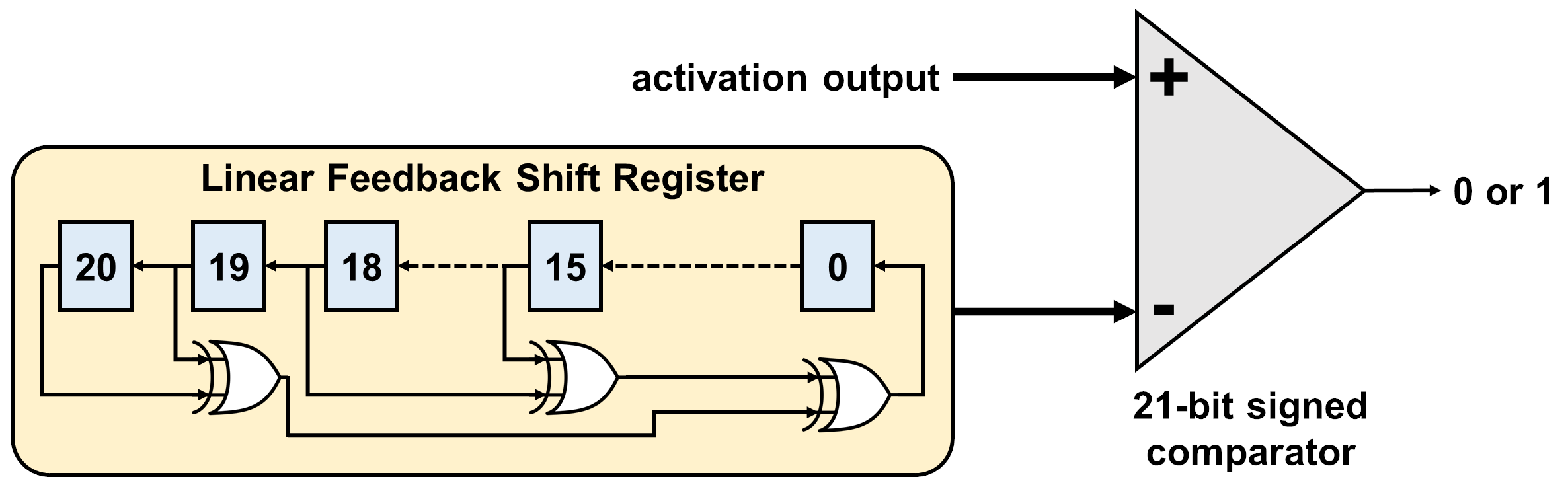}
\caption{Implementation of the p-bit stochasticity using linear feedback shift register (LFSR) and signed comparator.}
\label{fig:lfsr_comparator}
\end{figure}

The $rand( -1, +1)$ and $sgn( \, . \,)$ functions in step 6 of Algorithm \ref{algo:probabilistic_computing} represent the p-bit stochasticity. The  $rand( -1, +1)$ function is implemented using a 21-bit Fibonacci-style linear feedback shift register (LFSR) \cite{ward_lfsr_2007} with 1 sign bit and 20 fractional bits (consistent with the discussion in Section \ref{subsec:hyperparams}). The $sgn$ function is implemented using a signed comparator whose output is 0 or 1 (equivalent to -1 or +1 respectively), denoting the updated value of the p-bit in that sample.
The overall circuit diagram is shown in Fig. \ref{fig:lfsr_comparator}.

\subsection{Pseudo-Parallel p-Bit Update with Speculate-and-Select}
\label{subsec:pseudo_parallel_update}

\begin{figure}[!t]
\centering
\includegraphics[width=3.4in]{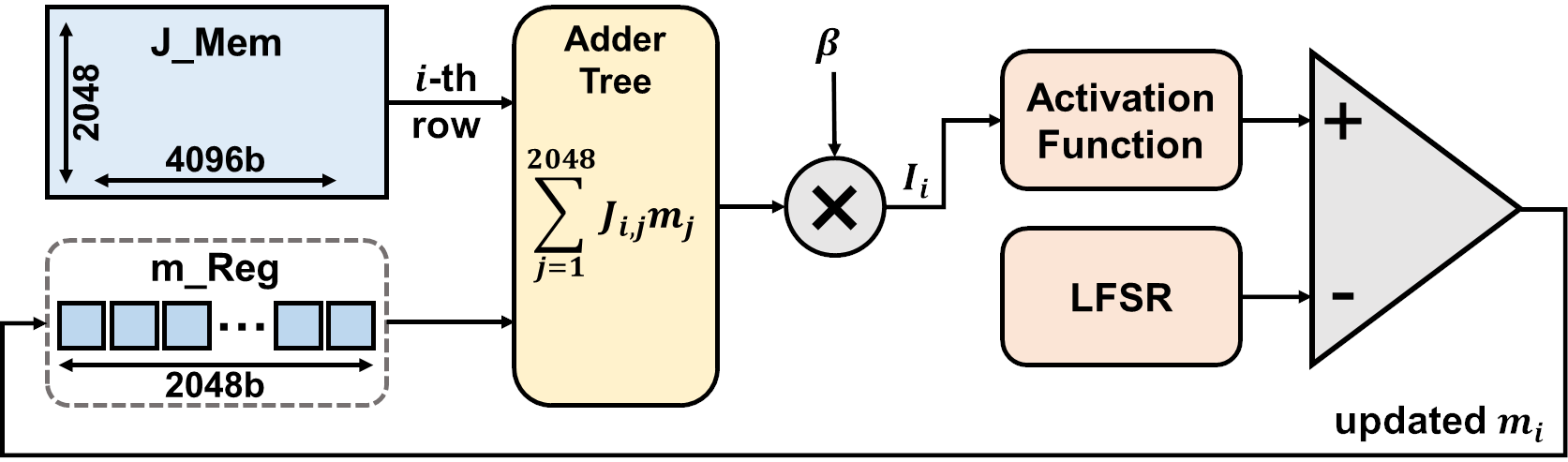}
\caption{Architecture of the baseline sequential p-bit update core.}
\label{fig:core_arch_sequential}
\end{figure}

Fig. \ref{fig:core_arch_sequential} shows a baseline architecture of the p-Bit Update Core which integrates the J\_Mem, m\_Reg, adder tree, $\beta$-multiplier, activation function, LFSR and comparator to update the p-bit state sequentially one p-bit at a time according to Algorithm \ref{algo:probabilistic_computing}. It takes $N_m + 1$ cycles to update all the $N_m$ p-bits in each sample, and hence requires $(N_m + 1) N_s$ cycles to complete $N_s$ samples and reach the final p-bit state. However, this sequential architecture limits the overall performance of the system as it can update only one p-bit per clock cycle.

\begin{figure}[!t]
\centering
\includegraphics[width=3.4in]{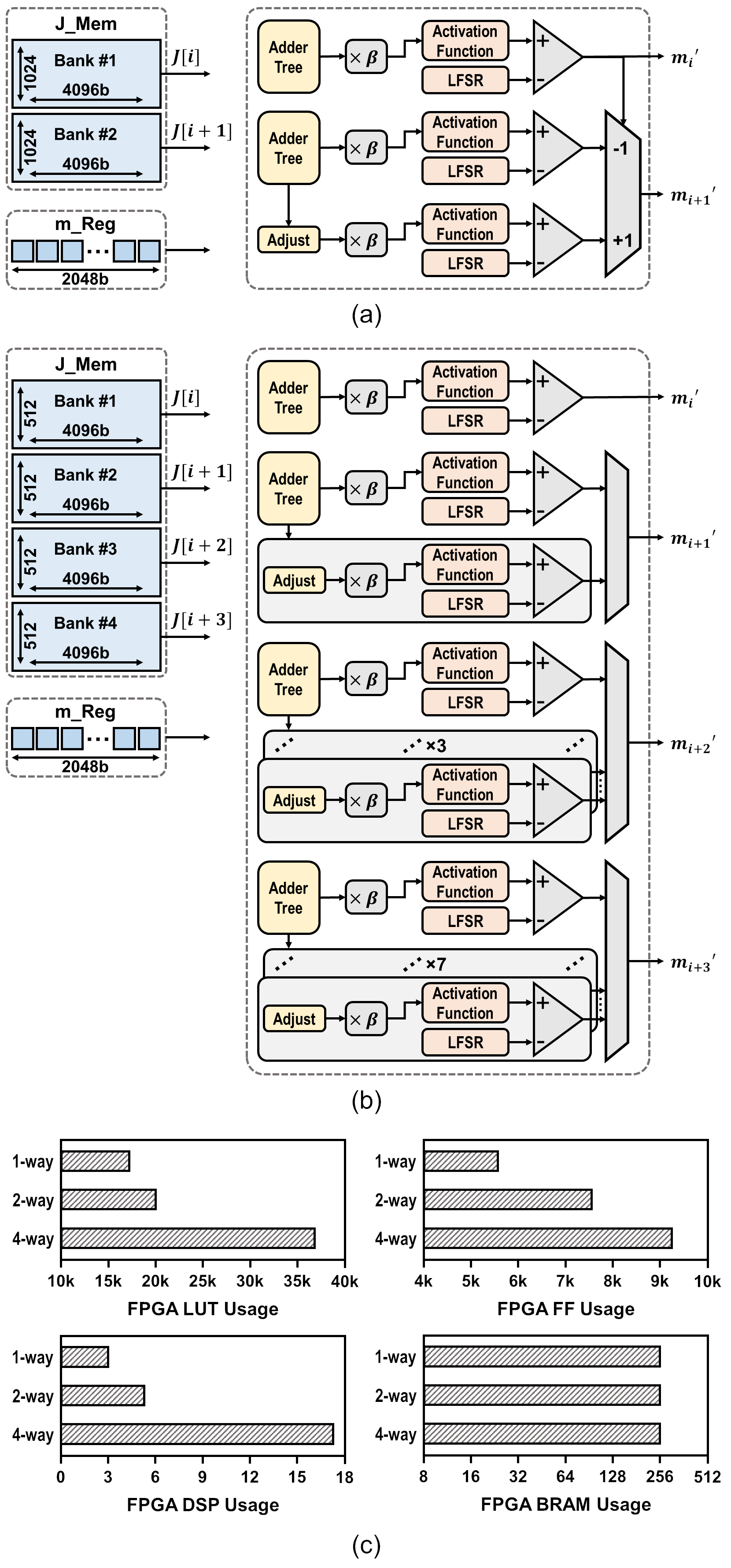}
\caption{Architectures of the proposed (a) 2-way and (b) 4-way pseudo-parallel p-bit update cores with speculate-and-select logic, and (c) comparison of FPGA resource utilization of 1-way, 2-way and 4-way architectures.}
\label{fig:core_arch_parallel}
\end{figure}

The sequential nature of the p-bit updates is inspired by Gibbs sampling which does not allow independently updating multiple p-bits in parallel. However, it is possible to speculatively compute all possible combinations of multiple updated p-bit values in parallel and then select the appropriate ones at the end.
This technique resembles carry-select logic in high-performance adders \cite{rabaey_chandrakasan, weste_harris, huang_tcas2_2023}.
For example, let us consider the computation of the updated value of $m_i$ as:
\[
m_i' = sgn ( \, rand( -1, +1) + tanh( \, \beta \times ( \, \sum_{j=1}^{N_m} \, J_{i,j} m_{j} \, ) \, ) \, )
\]
which has two possibilities: $m_i' = -1$ or $m_i' = +1$. Then, we can also simultaneously pre-compute the following two values:
\begin{align*}
m_{i+1}'|_{m_i' = -1} &= sgn ( \, rand( -1, +1) \,\, + \\
& tanh( \, \beta \times ( \, \sum_{1 \le j \le N_m, j \ne i} \, J_{i+1,j} m_{j} - J_{i+1,i} \, ) \, ) \, ) \\
m_{i+1}'|_{m_i' = +1} &= sgn ( \, rand( -1, +1) \,\, + \\
& tanh( \, \beta \times ( \, \sum_{1 \le j \le N_m, j \ne i} \, J_{i+1,j} m_{j} + J_{i+1,i} \, ) \, ) \, )
\end{align*}
which are the two possible updated values of $m_{i+1}$ if the updated value of $m_i$ is -1 and +1 respectively. All three values $m_i'$, $m_{i+1}'|_{m_i' = -1}$ and $m_{i+1}'|_{m_i' = +1}$ are computed in parallel, and then the correct $m_{i+1}'$ is selected as:
\[
m_{i+1}' = \begin{cases} 
    m_{i+1}'|_{m_i' = -1} & \text{if} \,\, m_i' = -1 \\
    m_{i+1}'|_{m_i' = +1} & \text{if} \,\, m_i' = +1
\end{cases}
\]
Therefore, the p-bits can be updated two at a time for $i \in \{1, 3, 5, \cdots , 2047\}$. Note that the dependency of p-bit updates is maintained, that is, Algorithm \ref{algo:probabilistic_computing} is still followed. Therefore, we refer to this technique as \textit{pseudo-parallel update} with \textit{speculate-and-select}. In particular, the above equations describe the \textit{2-way pseudo-parallel architecture} where 2 p-bits are updated per cycle, that is, overall $(\frac{N_m}{2} + 1) N_s$ cycles are required to complete $N_s$ samples. To enable the pseudo-parallel computation of two p-bit updates, the J\_Mem is split into two banks so that both the $i$-th and $(i+1)$-th rows of matrix \textbf{J} (denoted as $J[i]$ and $J[i+1]$ respectively) can be read simultaneously in the same cycle. The overall architecture of the 2-way pseudo-parallel p-Bit Update Core is shown in Fig. \ref{fig:core_arch_parallel}a.
Apart from J\_Mem and m\_Reg, it requires 2 instances of the adder tree, 3 instances each of the $\beta$-multiplier, the activation function, the LFSR and the comparator, along with additional control circuitry and multiplexors.
Note that the speculative update of $m_{i+1}$ requires only one adder tree whose output is then adjusted according to the different speculations, e.g., $( \, \sum_{1 \le j \le N_m, j \ne i} \, J_{i+1,j} m_{j} - J_{i+1,i} \, )$ can be calculated in one path for $m_{i+1}'|_{m_i' = -1}$, and it can be adjusted by simply adding $2J_{i+1,i}$ in the other path for $m_{i+1}'|_{m_i' = +1}$.
Although multiple instances of the $\beta$-multiplier are required, it is sufficient to have only one multiplier for computing the annealing schedule ($\beta = \beta \times \beta_{anneal-rate}$).
This idea can be further extended to a \textit{4-way pseudo-parallel architecture} where 4 p-bits are updated per cycle, that is, overall $(\frac{N_m}{4} + 1) N_s$ cycles are required to complete $N_s$ samples.
To enable the pseudo-parallel computation of four p-bit updates, the J\_Mem is split into four banks so that the $i$-th, $(i+1)$-th, $(i+2)$-th and $(i+3)$-th rows of matrix \textbf{J} (denoted as $J[i]$, $J[i+1]$, $J[i+2]$ and $J[i+3]$ respectively) can be read simultaneously in the same cycle. The p-bits are updated four at a time for $i \in \{1, 5, 9, \cdots , 2045\}$. Fig. \ref{fig:core_arch_parallel}b shows the overall architecture of the 4-way pseudo-parallel p-Bit Update Core.
Apart from J\_Mem, m\_Reg, control circuitry and multiplexors, it requires 4 instances of the adder tree and 15 instances each of the $\beta$-multiplier, the activation function, the LFSR and the comparator.
Fig. \ref{fig:core_arch_parallel}c compares the proposed 2-way and 4-way pseudo-parallel architectures with the baseline 1-way sequential architecture in terms of FPGA resource utilization (LUTs, FFs, DSPs and BRAMs).
All three implementations require the same amount of BRAM slices, but increasing the number of pseudo-parallel updates significantly increases the other resource requirements.
In general, a $k$-way pseudo-parallel p-bit update architecture will require $k$ instances of the adder tree, $2^k - 1$ instances each of the $\beta$-multiplier, the activation function, the LFSR and the comparator, along with J\_Mem (split into $k$ banks), m\_Reg, multiplexors and control circuitry.
The logic resource utilization increases exponentially with increasing $k$ and our proposed architecture can be scaled to support more pseudo-parallel updates, e.g., 8-way, 16-way, etc, based on resources available in the target FPGA.
\section{Implementation Results}
\label{sec:implementation}

\begin{figure}[!t]
\centering
\includegraphics[width=3.4in]{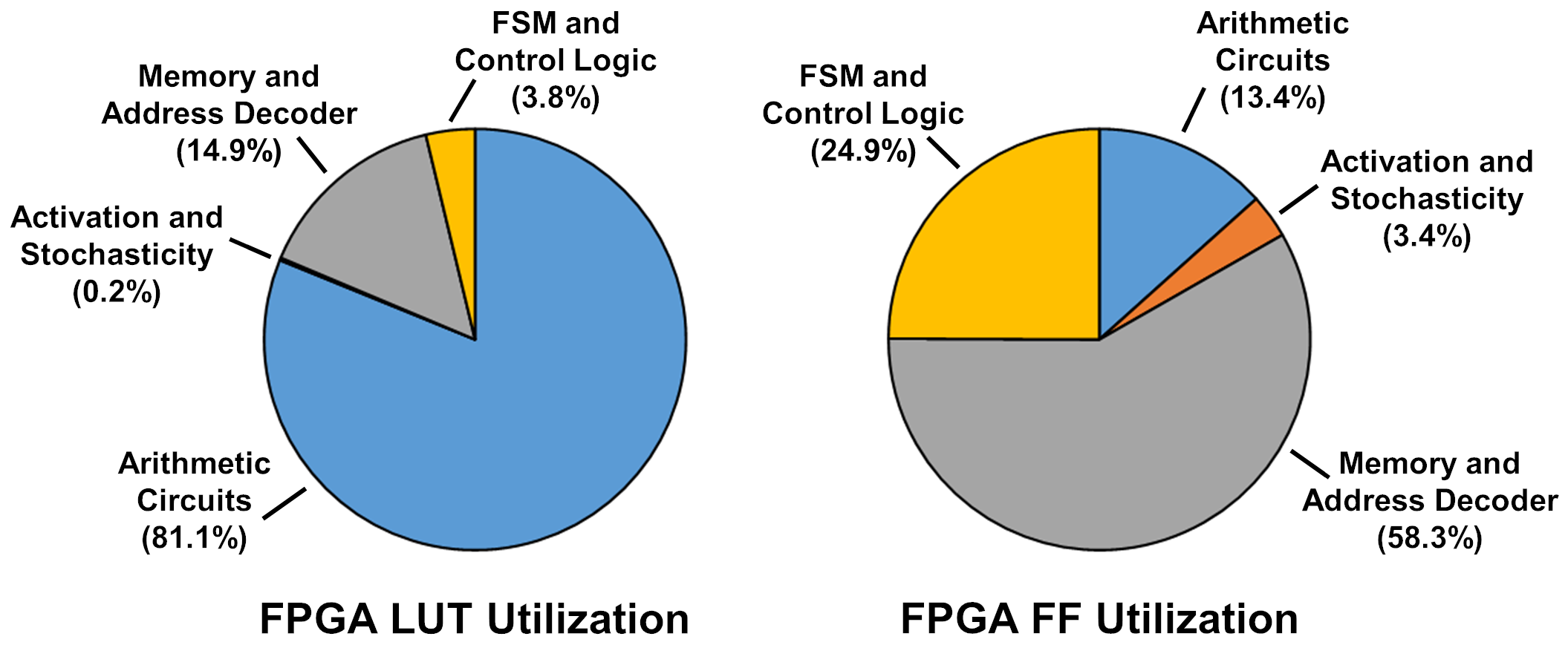}
\caption{Distribution of FPGA resource utilization of our proposed accelerator with 4-way pseudo-parallel p-bit update and speculate-and-select logic.}
\label{fig:fpga_utilization}
\end{figure}

\begin{figure}[!t]
\centering
\includegraphics[width=3.4in]{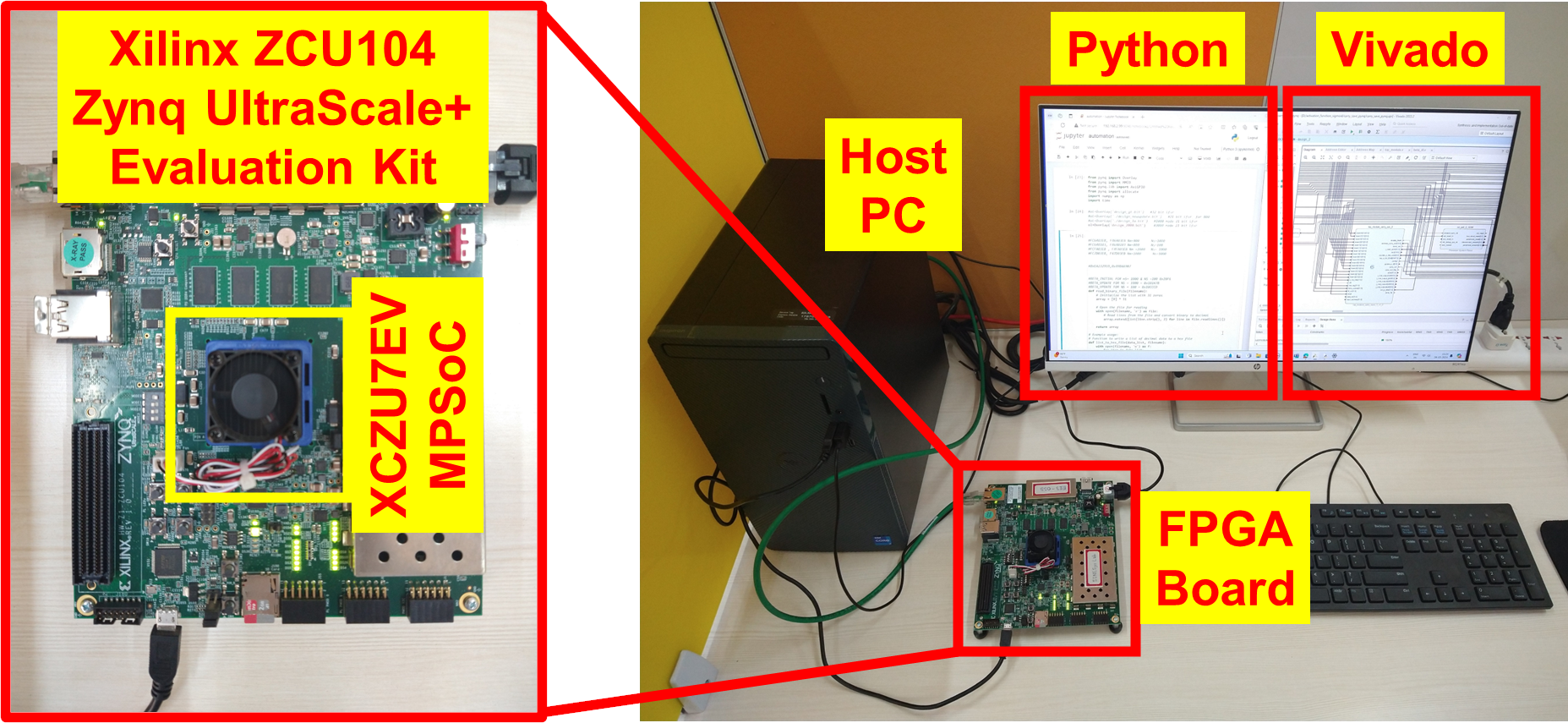}
\caption{Measurement setup with Zynq UltraScale+ ZCU104 FPGA board.}
\label{fig:fpga_demo_setup}
\end{figure}

We implement and validate our proposed accelerator on a Xilinx Zynq UltraScale+ MPSoC ZCU104 Evaluation Board with an XCZU7EV-2FFVC1156E device \cite{xilinx_ultrascale} using Verilog HDL and Xilinx Vivado Design Suite version ML 2022.2.
Its programmable logic (PL) contains 230k look-up tables (LUTs) and 461k flip-flops (FFs) in configurable logic blocks (CLBs), 312 Block RAMs (BRAMs) and 1,728 digital signal processing (DSP) slices, while its processing system (PS) consists of a quad-core ARM Cortex-A53 micro-processor.
Verilog HDL (Hardware Description Language) is used to design the hardware accelerator, and Xilinx Vivado Design Suite version ML 2022.2 is utilized for FPGA synthesis, implementation and simulation.
Our accelerator (with 4-way pseudo-parallel p-bit update) operates at a clock frequency of 100~MHz, and utilizes 37k LUTs, 9.5k FFs, 17 DSPs (equivalent to $\approx$ 7k LUTs \cite{zhang_fpga_2022}) and 256 BRAMs (total 8~Mb) in UltraScale+ FPGA.
The distribution of FPGA resource utilization
among different sub-modules of the accelerator
is shown in Fig. \ref{fig:fpga_utilization}.
Our experimental setup is shown in Fig. \ref{fig:fpga_demo_setup} with the FPGA board and the host PC running Python and Vivado interfaces.

\begin{figure}[!t]
\centering
\includegraphics[width=3.4in]{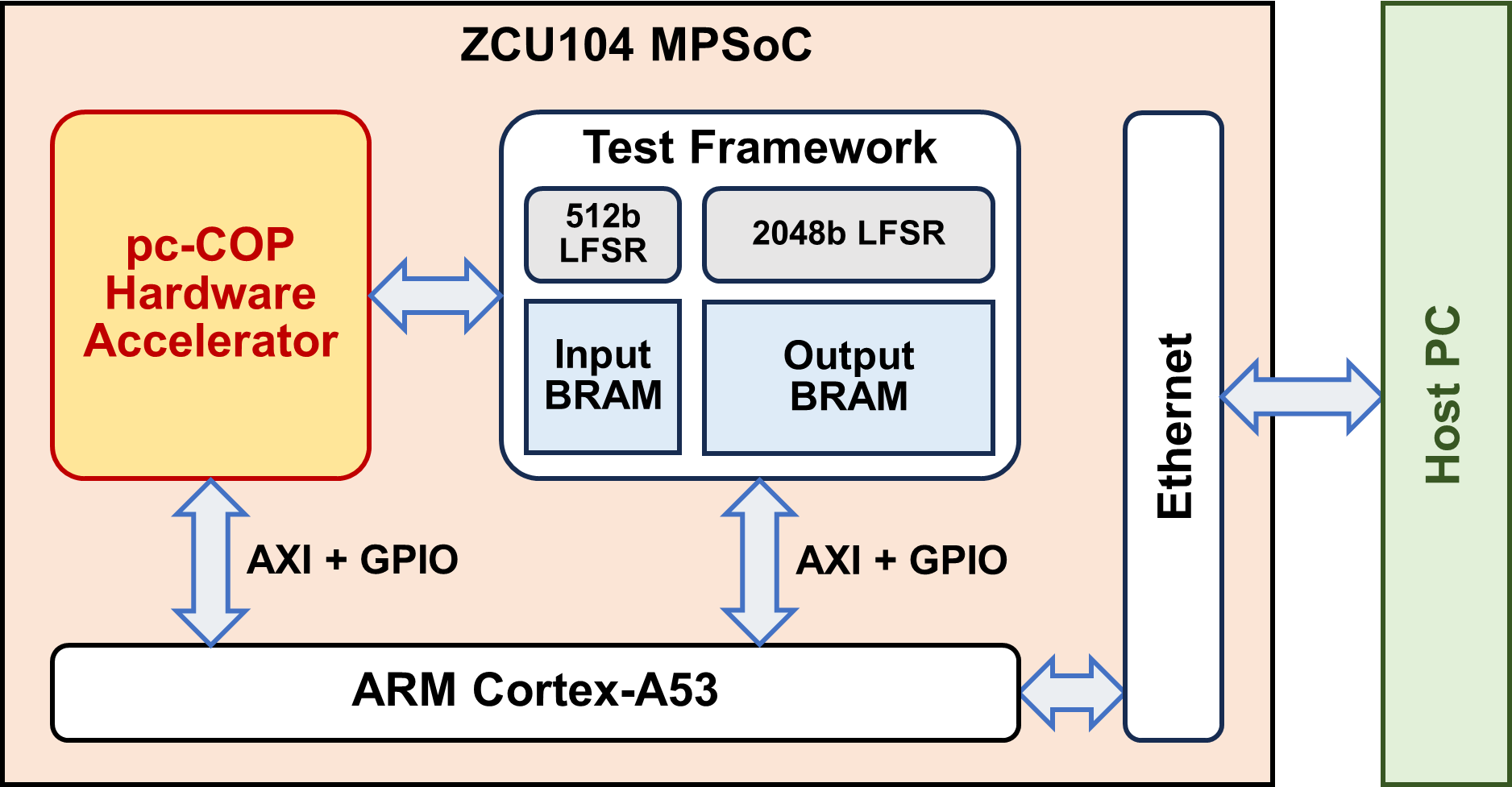}
\caption{Experimental validation framework consisting of proposed FPGA-based hardware accelerator interfaced with ARM processor in Zynq MPSoC.}
\label{fig:test_framework}
\end{figure}

Fig. \ref{fig:test_framework} provides an overview of our experimental validation framework consisting of the Zynq board and the host PC connected through Ethernet. The ARM processor in the Zynq PS is used to configure the J\_Mem with a problem-specific \textbf{J} matrix as well as provide the initial p-bit state, LFSR seed, annealing parameters and instruction through input ports of the accelerator implemented in the Zynq PL. A 2048-bit LFSR is used to generate the initial p-bit state, while another 512-bit LFSR is used to seed the pc-COP internal LFSRs.
Note that out of these 512 seed bits, 21, 63 and 315 bits are used for the 1-way, 2-way and 4-way designs with 1, 3 and 15 internal 21-bit LFSRs respectively.
Both the hardware accelerator and the test framework (containing an input BRAM, an output BRAM and the 2048-bit and 512-bit LFSRs) interface with the ARM processor through a 32-bit AXI interface and GPIOs.
The ARM processor is programmed using the open-source Python-based PYNQ software framework provided by Xilinx.

\begin{table}[!t]
\renewcommand{\arraystretch}{1.25}
\caption{pc-COP Measured Performance and Accuracy}
\label{table:gset_accuracy}
\centering
\begin{tabular}{|c|c|c|c|c|}
\hline
\textbf{No. of} & \textbf{Type} & \textbf{Benchmark} & \multicolumn{2}{c|}{\textbf{Average Accuracy} \textdagger} \\  \cline{4-5}
\textbf{Nodes} & \textbf{of Graphs} & \textbf{Graphs} & \textbf{\textbf{$N_s = 1000$}} & \textbf{\textbf{$N_s = 100$}} \\
\hline
\hline
800 & G-Set Random & G1 - G10 & 99.30\% & 97.33\% \\
\hline
800 & G-Set Toroidal & G11 - G13 & 95.65\% & 86.24\% \\
\hline
800 & G-Set Planar & G14 - G21 & 98.33\% & 94.46\% \\
\hline
1000 & G-Set Random & G43 - G47 & 99.63\% & 98.93\% \\
\hline
1000 & G-Set Planar & G51 - G54 & 99.05\% & 98.43\% \\
\hline
2000 & G-Set Random & G22 - G31 & 99.02\% & 97.22\% \\
\hline
2000 & G-Set Toroidal & G32 - G34 & 95.43\% & 91.12\% \\
\hline
2000 & G-Set Planar & G35 - G42 & 98.21\% & 96.72\% \\
\hline
2000 & Fully Connected & K2000 & 98.89\% & 97.99\% \\
\hline
\multicolumn{5}{l}{\footnotesize{\textdagger $\,\,\,$ measured results averaged over 1000 trials}}
\end{tabular}
\end{table}

\begin{figure}[!t]
\centering
\includegraphics[width=3.4in]{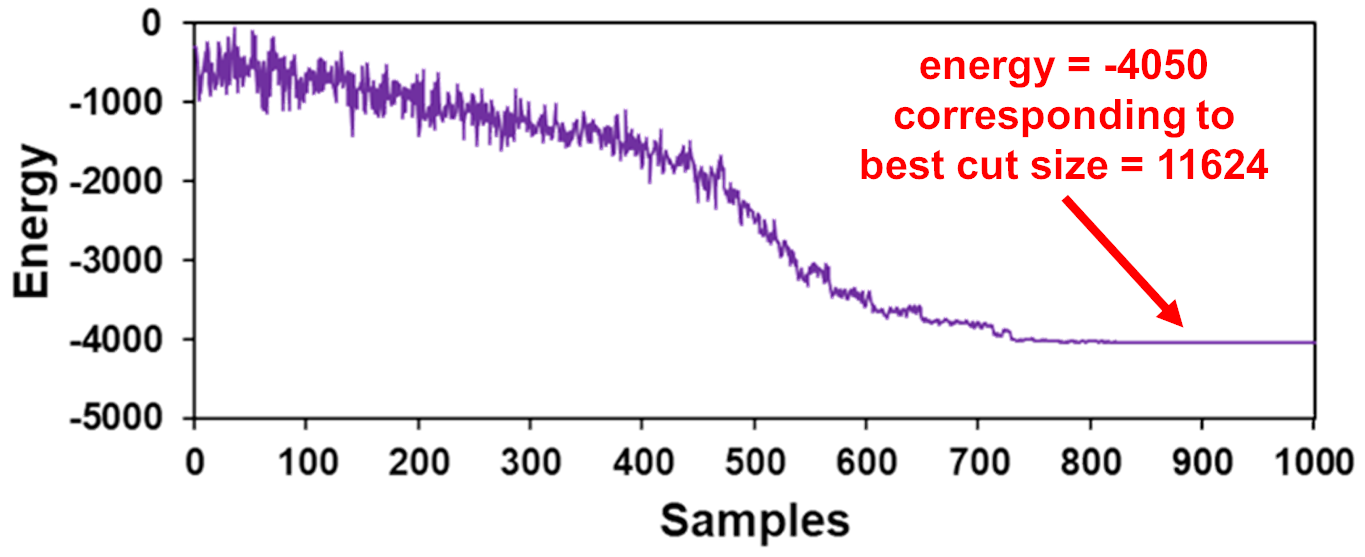}
\caption{Evolution of system energy and convergence towards solution for G1 benchmark with $N_s = 1000$, as measured from our experimental setup.}
\label{fig:energy_evolution}
\end{figure}

We use the standard G-Set max-cut benchmark graphs \cite{stanford_gset} with 800, 1000 and 2,000 nodes to evaluate the performance and accuracy of our design with the 4-way pseudo-parallel p-bit update explained in Section \ref{subsec:pseudo_parallel_update}.
We conduct 1000 trials for each G-Set benchmark for both $N_s = 1000$ and $N_s = 100$ (with the annealing hyper-parameters discussed in Section \ref{subsec:hyperparams}) to obtain a reasonable distribution of the accuracy of results (accuracy calculated relative to best known cut values from state-of-the-art \cite{toshiba_blog_2019}).
Each trial takes 2.01~ms, 2.51~ms and 5.01~ms respectively for $N_m$ = 800, 1000 and 2000 with $N_s = 1000$, and 201~$\mu$s, 251~$\mu$s and 501~$\mu$s respectively for $N_m$ = 800, 1000 and 2000 with $N_s = 100$.
pc-COP achieves an average accuracy of 98.49\% and 95.99\% across all the 51 evaluated G-Set graphs for $N_s = 1000$ and $N_s = 100$ respectively. Table \ref{table:gset_accuracy} shows the average accuracy for different graph sizes and types of graphs.
We note that pc-COP is able to reach near-99\% for most of the benchmarks, thus highlighting its potential for solving large-scale max-cut and other combinatorial optimization problems. Fig. \ref{fig:energy_evolution} shows the evolution of the system energy and convergence towards the best known cut size of 11624 (with final energy $E\{m\} = -4050$) for the G1 benchmark, as measured from our experimental setup.
We also evaluate max-cut performance with the K2000 benchmark \cite{onizawa_enhanced_2024} which is a fully-connected graph where all nodes are connected to each other with $\{ 0, \pm 1 \}$ weights. Across 1000 trials, we achieve average accuracy of 98.89\% and 97.99\% for $N_s = 1000$ and $N_s = 100$ respectively.

\clearpage

\begin{table*}[!t]
\renewcommand{\arraystretch}{1.19}
\caption{Max-Cut Benchmark Performance of our pc-COP Implementation on Xilinx UltraScale+ FPGA}
\label{table:benchmarks}
\centering
\begin{tabular}{|c|c|c|c|c|c|c|c|c|c|}
\hline
\textbf{} & \textbf{No.} & \textbf{No.} & \textbf{} & \textbf{Type} & \textbf{Best} & \multicolumn{4}{c|}{\textbf{pc-COP Measured Results} \textdagger} \\ \cline{7-10}
\textbf{Graph} & \textbf{of} & \textbf{of} & \textbf{Weights} & \textbf{of} & \textbf{Known} & \multicolumn{2}{c|}{\textbf{with $N_s = 1000$}} & \multicolumn{2}{c|}{\textbf{with $N_s = 100$}} \\ \cline{7-10}
\textbf{} & \textbf{Nodes} & \textbf{Edges} & \textbf{} & \textbf{Graph} & \textbf{Cut Value} & \textbf{$\,$ Best Cut $\,$} & \textbf{Accuracy} & \textbf{$\,$ Best Cut $\,$} & \textbf{Accuracy} \\
\hline
\hline
G1 & 800 & 19176 & $\{0, 1\}$ & Random & 11624 & 11624&99.75 &11585 & 99.08\\
\hline
G2 & 800 & 19176 & $\{0, 1\}$ & Random & 11620 &11620 &99.77 & 11595& 99.16\\
\hline
G3 & 800 & 19176 & $\{0, 1\}$ & Random & 11622 &11622 &99.80 &11583 &99.15 \\
\hline
G4 & 800 & 19176 & $\{0, 1\}$ & Random & 11646 &11646 &99.84 & 11594&99.10 \\
\hline
G5 & 800 & 19176 & $\{0, 1\}$ & Random & 11631 &11631 &99.83 &11597 &99.19 \\
\hline
G6 & 800 & 19176 & $\{0, \pm 1\}$ & Random & 2178 &2178 & 99.05&2158 &95.48 \\
\hline
G7 & 800 & 19176 & $\{0, \pm 1\}$ & Random & 2006 &2006 &98.69 & 1965&95.27 \\
\hline
G8 & 800 & 19176 & $\{0, \pm 1\}$ & Random & 2005 & 2005&98.89 & 1978& 96.10\\
\hline
G9 & 800 & 19176 & $\{0, \pm 1\}$ & Random & 2054 & 2053&98.71 &2017 & 95.38\\
\hline
G10 & 800 & 19176 & $\{0, \pm 1\}$ & Random & 2000 & 2000&98.68 &1967 &95.43 \\
\hline
G11 & 800 & 1600 & $\{0, \pm 1\}$ & Toroidal & 564 & 558&95.98 &520 & 86.25\\
\hline
G12 & 800 & 1600 & $\{0, \pm 1\}$ & Toroidal & 556 & 546& 95.39&522 & 86.25\\
\hline
G13 & 800 & 1600 & $\{0, \pm 1\}$ & Toroidal & 582 & 576&95.60 &538 &86.22 \\
\hline
G14 & 800 & 4694 & $\{0, 1\}$ & Planar & 3064 &3053 &99.09 & 3003&96.94 \\
\hline
G15 & 800 & 4661 & $\{0, 1\}$ & Planar & 3050 &3039 &99.00 &2984 &96.79 \\
\hline
G16 & 800 & 4672 & $\{0, 1\}$ & Planar & 3052 &3041 & 99.03&2982 &96.90 \\
\hline
G17 & 800 & 4667 & $\{0, 1\}$ & Planar & 3047 & 3034& 99.00 &2982 &96.81 \\
\hline
G18 & 800 & 4694 & $\{0, \pm 1\}$ & Planar & 992 &989 &97.71 &966 &92.60 \\
\hline
G19 & 800 & 4661 & $\{0, \pm 1\}$ & Planar & 906 &904 &97.41 &864 &91.76 \\
\hline
G20 & 800 & 4672 & $\{0, \pm 1\}$ & Planar & 941 &941 & 97.90&910 &91.97 \\
\hline
G21 & 800 & 4667 & $\{0, \pm 1\}$ & Planar & 931 & 930& 97.52&891 &91.92 \\
\hline
\hline
G43 & 1000 & 9990 & $\{0, 1\}$ & Random & 6660 & 6660 & 99.61& 6653&98.89 \\
\hline
G44 & 1000 & 9990 & $\{0, 1\}$ & Random & 6650 & 6648&99.67 &6627 &98.95 \\
\hline
G45 & 1000 & 9990 & $\{0, 1\}$ & Random & 6654 &6653 &99.63 &6633 &98.92 \\
\hline
G46 & 1000 & 9990 & $\{0, 1\}$ & Random & 6649 &6646 &99.64 &6630 &98.98 \\
\hline
G47 & 1000 & 9990 & $\{0, 1\}$ & Random & 6657 &6655 & 99.64&6643 & 98.95\\
\hline
G51 & 1000 & 5909 & $\{0, 1\}$ & Planar & 3848 &3830 & 99.04&3819 & 98.43\\
\hline
G52 & 1000 & 5916 & $\{0, 1\}$ & Planar & 3851 &3834&99.06 &3818 &98.49 \\
\hline
G53 & 1000 & 5914 & $\{0, 1\}$ & Planar & 3850 & 3834&99.10 & 3819& 98.47\\
\hline
G54 & 1000 & 5916 & $\{0, 1\}$ & Planar & 3852 & 3840&99.01 &3816 & 98.35\\
\hline
\hline
G22 & 2000 & 19990 & $\{0, 1\}$ & Random & 13359 &13352 &99.57 &13281 &98.78 \\
\hline
G23 & 2000 & 19990 & $\{0, 1\}$ & Random & 13344 &13331 &99.67 &13281 & 98.93\\
\hline
G24 & 2000 & 19990 & $\{0, 1\}$ & Random & 13337 &13326 & 99.63& 13260&98.91 \\
\hline
G25 & 2000 & 19990 & $\{0, 1\}$ & Random & 13340 &13335 & 99.65&13262 &98.91 \\
\hline
G26 & 2000 & 19990 & $\{0, 1\}$ & Random & 13328 &13317 &99.64 &13257 &98.95 \\
\hline
G27 & 2000 & 19990 & $\{0, \pm 1\}$ & Random & 3341 & 3330& 98.38&3282 &95.51 \\
\hline
G28 & 2000 & 19990 & $\{0, \pm 1\}$ & Random & 3298 &3289 &98.45 &3081 &95.48 \\
\hline
G29 & 2000 & 19990 & $\{0, \pm 1\}$ & Random & 3405 &3394 & 98.24& 3326&95.54 \\
\hline
G30 & 2000 & 19990 & $\{0, \pm 1\}$ & Random & 3413 &3403 &98.52 &3335 &95.61 \\
\hline
G31 & 2000 & 19990 & $\{0, \pm 1\}$ & Random & 3310 &3297 &98.45 & 3246&95.60 \\
\hline
G32 & 2000 & 4000 & $\{0, \pm 1\}$ & Toroidal & 1410 & 1370&95.23 &1322 &90.97 \\
\hline
G33 & 2000 & 4000 & $\{0, \pm 1\}$ & Toroidal & 1382 &1348 & 95.41&1302 &91.06 \\
\hline
G34 & 2000 & 4000 & $\{0, \pm 1\}$ & Toroidal & 1384 &1348 & 95.65& 1308& 91.34\\
\hline
G35 & 2000 & 11778 & $\{0, 1\}$ & Planar & 7687 &7644 &98.99 &7592 &98.37 \\
\hline
G36 & 2000 & 11766 & $\{0, 1\}$ & Planar & 7680 & 7633& 98.99& 7592& 98.39\\
\hline
G37 & 2000 & 11785 & $\{0, 1\}$ & Planar & 7691 &7652 &98.97 &7608 & 98.36\\
\hline
G38 & 2000 & 11779 & $\{0, 1\}$ & Planar & 7688 &7657 & 99.01&7602 & 98.35\\
\hline
G39 & 2000 & 11778 & $\{0, \pm 1\}$ & Planar & 2408 & 2392&97.51 & 2346& 95.32\\
\hline
G40 & 2000 & 11766 & $\{0, \pm 1\}$ & Planar & 2400 & 2382&97.37 &2339 &95.00 \\
\hline
G41 & 2000 & 11785 & $\{0, \pm 1\}$ & Planar & 2405 & 2383&97.31 &2341 &94.81 \\
\hline
G42 & 2000 & 11779 & $\{0, \pm 1\}$ & Planar & 2481 &2455 &97.53 &2424 &95.17 \\
\hline
\hline
K2000 & 2000 & 1999000 & $\{0, \pm 1\}$ & Full & 33337 &33101 &98.89 & 32670&97.99 \\
\hline
\multicolumn{10}{l}{\footnotesize{\textdagger $\,\,\,$ measured results for best cut value and average accuracy over 1000 trials for both $N_s = 1000$ and $N_s = 100$}}
\end{tabular}
\end{table*}

\clearpage

\begin{table*}[!t]
\renewcommand{\arraystretch}{1.2}
\caption{Comparison of pc-COP with State-of-the-Art FPGA-Based G-Set Max-Cut COP Hardware Accelerators}
\label{table:comparison}
\centering
\begin{tabular}{|l|c|c|c|c|c|c|c|c|c|}
\hline
\textbf{Design} & \textbf{Type} & \textbf{Tech} & \textbf{No. of} & \textbf{Connection} & \textbf{Weight} & \textbf{Resource} & \textbf{Op.} & \textbf{G-Set Avg.} & \textbf{Time to} \\
& & & \textbf{Nodes} & \textbf{Topology} & \textbf{Precision} & \textbf{Utilization} & \textbf{Freq.} & \textbf{Accuracy} & \textbf{Solution} \\
\hline
\hline
\multirow{2}{*}{\cite{cook_gpu_2019}} & Digital & 22nm & 800 - & Fully & \multirow{2}{*}{2 bits} & 32 Xeon cores & 2.3 & \multirow{2}{*}{95.61\%} & 170~ms - \\
& Annealing & CPU & 20000 & Connected & & + 72~GB DRAM & GHz & & 19.89~s \\
\hline
\multirow{2}{*}{\cite{cook_gpu_2019}} & Digital & 28nm & 800 - & Fully & \multirow{2}{*}{2 bits} & 2880 CUDA cores & 745 & \multirow{2}{*}{95.61\%} & 110~ms - \\
& Annealing & GPU & 20000 & Connected & & + 12~GB DRAM & MHz & & 390~ms \\
\hline
\multirow{2}{*}{\cite{inagaki_cim_2016}} & Coherent & \multirow{2}{*}{Optics} & \multirow{2}{*}{2000} & Fully & \multirow{2}{*}{1 bit} & \multirow{2}{*}{$-$} & \multirow{2}{*}{$-$} & \multirow{2}{*}{97.92\%} & \multirow{2}{*}{5~ms} \\
& Ising Machine & & & Connected & & & & & \\
\hline
\multirow{2}{*}{\cite{huang_tcas2_2023}} & Digital & 16nm & \multirow{2}{*}{1024} & Fully & \multirow{2}{*}{4 bits} & 40k LUTs + 12k FFs & 100 & \multirow{2}{*}{99.07\%} & 373~$\mu$s - \\
& Annealing & FPGA & & Connected & & + 4~Mb BRAM & MHz & & 5.38~ms \\
\hline
\multirow{2}{*}{\cite{huang_iscas_2023}} & Digital & 16nm & \multirow{2}{*}{1024} & Fully & \multirow{2}{*}{4 bits} & 75k LUTs + 12k FFs & 100 & \multirow{2}{*}{99.19\%} & 186~$\mu$s - \\
& Annealing & FPGA & & Connected & & + 4~Mb BRAM & MHz & & 1.35~ms \\
\hline
\multirow{2}{*}{\cite{waidyasooriya_temporal_2022}} & Digital & 20nm & \multirow{2}{*}{4096} & Fully & \multirow{2}{*}{2 bits} & \multirow{2}{*}{$-$} & \multirow{2}{*}{$-$} & \multirow{2}{*}{98.50\%} & 5~ms - \\
& Annealing & FPGA & & Connected & & & & & 25~ms \\
\hline
\multirow{2}{*}{\cite{zhang_tempering_2024}} & Parallel & 16nm & 1024 & Fully & \multirow{2}{*}{2 bits} & 99k LUTs + 74k FFs & 200 & \multirow{2}{*}{99.43\%} & 0.5~ms - \\
& Tempering & FPGA & ($\times8$ replicas) & Connected & & + 7.125~Mb BRAM & MHz & & 1~ms \\
\hline
\multirow{2}{*}{\cite{khan_maxcut_2022}} & Probabilistic & 14nm & 800 - & Fully & \multirow{2}{*}{2 bits} &\multirow{2}{*} {2 Core-i7 cores} & 2.5 & \multirow{2}{*}{$\approx$ 97.00\%}  & \multirow{2}{*}{$-$} \\
& Computing & CPU & 3000 & Connected & & & GHz & &  \\
\hline
& & & \textbf{\multirow{6}{*} {2048}} & & \textbf{\multirow{6}{*}{2 bits}} & & & \multicolumn{2}{c|}{\textbf{for $N_s = 1000$}} \\ \cline{9-10}
& & & & & & \textbf{37k LUTs + 9.5k FFs} & & \textbf{\multirow{2}{*}{98.49\%}} & \textbf{2.01~ms -} \\
\textbf{This} & \textbf{Probabilistic} & \textbf{16nm} & & \textbf{Fully} & & \textbf{+ 17 DSPs ($\approx$ 7k LUTs \textdagger)} & \textbf{100} & & \textbf{5.01~ms}  \\ \cline{9-10}
\textbf{Work} & \textbf{Computing} & \textbf{FPGA} & & \textbf{Connected} & & \textbf{+ 8~Mb BRAM} & \textbf{MHz} & \multicolumn{2}{c|}{\textbf{for $N_s = 100$}}  \\ \cline{9-10}
& & & & & & & & \textbf{\multirow{2}{*}{95.99\%}} & \textbf{201~$\mu$s -} \\
& & & & & & & & & \textbf{501~$\mu$s} \\
\hline
\multicolumn{10}{l}{\footnotesize{\textdagger $\,\,\,$ 1 DSP is equivalent to $\approx$ 51.2 logic slices \cite{zhang_fpga_2022} and 1 logic slice contains 8 LUTs in UltraScale+ FPGA \cite{xilinx_ultrascale}}}
\end{tabular}
\end{table*}

The detailed experimental results from our FPGA-based hardware implementation are presented in Table \ref{table:benchmarks}, with the best cut value and average accuracy (across 1000 trials) obtained for each benchmark graph.
Table \ref{table:comparison} compares our design with previous work on FPGA-based hardware accelerators demonstrating max-cut with G-Set benchmarks. Most of the previous work are digital annealers and Ising computers implemented using CPU and GPU \cite{cook_gpu_2019}, optics \cite{inagaki_cim_2016} and FPGA \cite{huang_tcas2_2023, huang_iscas_2023, waidyasooriya_temporal_2022, zhang_tempering_2024}. While there are many other implementations of FPGA-based and ASIC-based digital annealers in recent literature \cite{yoshimura_architecture_2020, kulkarni_isingcim_2022, kim_ctleising_2024}, we only include those which have demonstrated G-Set benchmarks for fair comparison. \cite{khan_maxcut_2022} is a CPU-based demonstration of G-Set max-cut with probabilistic computing.
Compared to previous CPU-based and GPU-based implementations, we achieve 3 orders of magnitude speedup while maintaining similar accuracy levels.
Compared to previous FPGA-based digital annealer implementations, we achieve reasonably comparable performance and accuracy with the new probabilistic computing paradigm while having lower FPGA resource utilization.
This clearly demonstrates that hardware-accelerated probabilistic computing is an excellent candidate for realizing efficient and large-scale combinatorial optimization problem solvers.
\section{Conclusion}
\label{sec:conclusion}

Probabilistic computing is an emerging quantum-inspired computing paradigm capable of solving various classes of computationally hard problems such as combinatorial optimization.
In this work, we present pc-COP, an efficient and configurable probabilistic computing hardware accelerator with 2048 fully connected p-bits implemented on Xilinx UltraScale+ FPGA and demonstrate the standard G-Set graph maximum cut benchmarks.
Our efficient logarithmic adder tree design for sum-of-products computation reduces critical path delay. We efficiently approximate the activation function and tune the precision of the annealing schedule to save logic resources. Finally, we propose a pseudo-parallel p-bit update architecture with speculate-and-select logic which improves overall performance by $4 \times$ compared to the traditional sequential p-bit update. We achieve near-99\% average accuracy across various G-Set max-cut benchmarks with 800, 1000 and 2000 nodes.
Our experimental results demonstrate that FPGA-based probabilistic computing hardware accelerators are promising practical systems for efficiently solving large-scale combinatorial optimization problems.
Future extensions of our work will explore larger designs with high-precision interaction coefficients, efficient memory architectures exploiting graph sparsity, problem-specific tuning of hyper-parameters and extensions to other problems such as traveling salesman and Boolean satisfiability.

\section*{Acknowledgment}

This work was supported in part by a seed grant from the Indian Institute of Science and in part by a Ph.D. Scholarship from the Ministry of Education, Government of India. The authors would like to thank Yashash Jain for helpful technical discussions and Dr. Shantharam Kalipatnapu for helping set up the PYNQ interface.


\bibliographystyle{IEEEtran}
\bibliography{references}

\begin{thebibliography}{10}
\providecommand{\url}[1]{#1}
\csname url@samestyle\endcsname
\providecommand{\newblock}{\relax}
\providecommand{\bibinfo}[2]{#2}
\providecommand{\BIBentrySTDinterwordspacing}{\spaceskip=0pt\relax}
\providecommand{\BIBentryALTinterwordstretchfactor}{4}
\providecommand{\BIBentryALTinterwordspacing}{\spaceskip=\fontdimen2\font plus
\BIBentryALTinterwordstretchfactor\fontdimen3\font minus \fontdimen4\font\relax}
\providecommand{\BIBforeignlanguage}[2]{{%
\expandafter\ifx\csname l@#1\endcsname\relax
\typeout{** WARNING: IEEEtran.bst: No hyphenation pattern has been}%
\typeout{** loaded for the language `#1'. Using the pattern for}%
\typeout{** the default language instead.}%
\else
\language=\csname l@#1\endcsname
\fi
#2}}
\providecommand{\BIBdecl}{\relax}
\BIBdecl

\bibitem{feynman_simulating_1982}
R.~P. Feynman, ``{Simulating Physics with Computers},'' \emph{International Journal of Theoretical Physics}, vol.~21, no. 6/7, 1982.

\bibitem{nielsen_chuang}
M.~A. {Nielsen} and I.~L. {Chuang}, \emph{{Quantum Computation and Quantum Information}}.\hskip 1em plus 0.5em minus 0.4em\relax Cambridge University Press, 2010.

\bibitem{shor_quantum_1997}
P.~W. {Shor}, ``{Polynomial-Time Algorithms for Prime Factorization and Discrete Logarithms on a Quantum Computer},'' \emph{SIAM Journal of Computing}, vol.~26, no.~5, pp. 1484--1509, Oct. 1997.

\bibitem{google_supremacy_2019}
F.~Arute \emph{et~al.}, ``{Quantum Supremacy using a Programmable Superconducting Processor},'' \emph{Nature}, vol. 574, no. 7779, pp. 505--510, 2019.

\bibitem{camsari_pbits_2017}
K.~Y. Camsari, R.~Faria, B.~M. Sutton, and S.~Datta, ``{Stochastic $p$-Bits for Invertible Logic},'' \emph{Phys. Rev. X}, vol.~7, p. 031014, Jul. 2017.

\bibitem{smithson_stochastic_2019}
S.~C. Smithson \emph{et~al.}, ``{Efficient CMOS Invertible Logic Using Stochastic Computing},'' \emph{IEEE Transactions on Circuits and Systems I: Regular Papers}, vol.~66, no.~6, pp. 2263--2274, Jan. 2019.

\bibitem{cook_gpu_2019}
C.~Cook \emph{et~al.}, ``{GPU-Based Ising Computing for Solving Max-Cut Combinatorial Optimization Problems},'' \emph{Integration}, vol.~69, pp. 335--344, 2019.

\bibitem{jung_annealing_2023}
H.~Jung \emph{et~al.}, ``{A Quantum-Inspired Probabilistic Prime Factorization based on Virtually Connected Boltzmann Machine and Probabilistic Annealing},'' \emph{Nature Scientific Reports}, vol.~13, no.~1, p. 16186, 2023.

\bibitem{zhang_tempering_2024}
Y.~Zhang \emph{et~al.}, ``{A Parallel Tempering Processing Architecture with Multi-Spin Update for Fully-Connected Ising Models},'' in \emph{IEEE Design, Automation \& Test in Europe Conference \& Exhibition (DATE)}, 2024, pp. 1--6.

\bibitem{camsari_dialogue_2021}
K.~Y. Camsari and S.~Datta, ``{Dialogue Concerning the Two Chief Computing Systems: Imagine Yourself on a Flight Talking to an Engineer About a Scheme that Straddles Classical and Quantum},'' \emph{IEEE Spectrum}, vol.~58, no.~4, pp. 30--35, Apr. 2021.

\bibitem{chowdhury_fullstack_2023}
S.~Chowdhury \emph{et~al.}, ``{A Full-Stack View of Probabilistic Computing with p-Bits: Devices, Architectures and Algorithms},'' \emph{IEEE Journal on Exploratory Solid-State Computational Devices and Circuits}, vol.~9, no.~1, pp. 1--11, Mar. 2023.

\bibitem{pervaiz_emulation_2017}
A.~Z. Pervaiz \emph{et~al.}, ``{Hardware Emulation of Stochastic p-Bits for Invertible Logic},'' \emph{Nature Scientific Reports}, vol.~7, no. 10994, Sep. 2017.

\bibitem{khan_maxcut_2022}
M.~Khan and O.~Hassan, ``{Benchmarking of Probabilistic-Bit Based Algorithm for Max-Cut Problem},'' in \emph{International Conference on Electrical and Computer Engineering (ICECE)}.\hskip 1em plus 0.5em minus 0.4em\relax IEEE, 2022, pp. 453--456.

\bibitem{onizawa_fastconv_2023}
N.~Onizawa \emph{et~al.}, ``{Fast-Converging Simulated Annealing for Ising Models Based on Integral Stochastic Computing},'' \emph{IEEE Transactions on Neural Networks and Learning Systems}, vol.~34, no.~12, pp. 10\,999--11\,005, Dec. 2023.

\bibitem{onizawa_enhanced_2024}
N.~Onizawa and T.~Hanyu, ``{Enhanced Convergence in p-Bit Based Simulated Annealing with Partial Deactivation for Large-Scale Combinatorial Optimization Problems},'' \emph{Nature Scientific Reports}, vol.~14, no.~1, p. 1339, 2024.

\bibitem{pervaiz_fpga_2019}
A.~Z. Pervaiz \emph{et~al.}, ``{Weighted $p$-Bits for FPGA Implementation of Probabilistic Circuits},'' \emph{IEEE Transactions on Neural Networks and Learning Systems}, vol.~30, no.~6, pp. 1920--1926, Oct. 2019.

\bibitem{borders_factorization_2019}
W.~A. Borders, A.~Z. Pervaiz, S.~Fukami, K.~Y. Camsari, H.~Ohno, and S.~Datta, ``{Integer Factorization using Stochastic Magnetic Tunnel Junctions},'' \emph{Nature}, vol. 573, no. 7774, pp. 390--393, Sep. 2019.

\bibitem{liu_probabilistic_2022}
Y.~Liu \emph{et~al.}, ``{Probabilistic Circuit Implementation Based on p-Bits using the Intrinsic Random Property of RRAM and p-Bit Multiplexing Strategy},'' \emph{Micromachines}, vol.~13, no.~6, p. 924, Jun. 2022.

\bibitem{luo_ferroelectric_2023}
S.~Luo, Y.~He, B.~Cai, X.~Gong, and G.~Liang, ``{Probabilistic-Bits Based on Ferroelectric Field-Effect Transistors for Probabilistic Computing},'' \emph{IEEE Electron Device Letters}, vol.~44, no.~8, pp. 1356--1359, 2023.

\bibitem{heo_oscillation_2023}
S.~Heo \emph{et~al.}, ``{Experimental Demonstration of Probabilistic-Bit (p-bit) Utilizing Stochastic Oscillation of Threshold Switch Device},'' in \emph{IEEE Symposium on VLSI Technology and Circuits (VLSI)}, 2023, pp. 1--2.

\bibitem{lucas_ising_2014}
A.~Lucas, ``{Ising Formulations of Many NP Problems},'' \emph{Frontiers in Physics}, vol.~2, p. 74887, 2014.

\bibitem{mohseni_ising_2022}
N.~Mohseni, P.~L. McMahon, and T.~Byrnes, ``{Ising Machines as Hardware Solvers of Combinatorial Optimization Problems},'' \emph{Nature Reviews Physics}, vol.~4, no.~6, pp. 363--379, 2022.

\bibitem{xilinx_ultrascale}
{Xilinx Inc.}, ``{UltraScale Architecture: Staying a Generation Ahead with an Extra Node of Value},'' \url{https://www.xilinx.com/products/technology/ultrascale.html}.

\bibitem{toshiba_blog_2019}
Y.~Matsuda, ``{Benchmarking the MAX-CUT Problem on the Simulated Bifurcation Machine},'' Medium, 2019, \url{https://medium.com/toshiba-sbm/benchmarking-the-max-cut-problem-on-the-simulated-bifurcation-machine-e26e1127c0b0}.

\bibitem{stanford_gset}
{Stanford University}, ``{G-Set Graph Dataset},'' \url{https://web.stanford.edu/\%7Eyyye/yyye/Gset}.

\bibitem{camsari_pbits_2019}
K.~Y. Camsari, B.~M. Sutton, and S.~Datta, ``{$p$-Bits for Probabilistic Spin Logic},'' \emph{Applied Physics Reviews}, vol.~6, no.~1, Mar. 2019.

\bibitem{jain_tyche_2023}
Y.~Jain and U.~Banerjee, ``{Tyche: A Compact and Configurable Accelerator for Scalable Probabilistic Computing on FPGA},'' in \emph{IEEE High Performance Extreme Computing Conference (HPEC)}, 2023, pp. 1--7.

\bibitem{sutton_autonomous_2020}
B.~Sutton, R.~Faria, L.~A. Ghantasala, R.~Jaiswal, K.~Y. Camsari, and S.~Datta, ``{Autonomous Probabilistic Coprocessing With Petaflips per Second},'' \emph{IEEE Access}, vol.~8, pp. 157\,238--157\,252, Aug. 2020.

\bibitem{aadit_computing_2021}
N.~A. Aadit \emph{et~al.}, ``{Computing with Invertible Logic: Combinatorial Optimization with Probabilistic Bits},'' in \emph{IEEE International Electron Devices Meeting (IEDM)}, Dec. 2021, pp. 40--43.

\bibitem{kaiser_probabilistic_2021}
J.~Kaiser and S.~Datta, ``{Probabilistic Computing with p-Bits},'' \emph{Applied Physics Letters}, vol. 119, no.~15, Oct. 2021.

\bibitem{aadit_massively_2022}
N.~A. Aadit \emph{et~al.}, ``{Massively Parallel Probabilistic Computing with Sparse Ising Machines},'' \emph{Nature Electronics}, vol.~5, no.~7, pp. 460--468, Jul. 2022.

\bibitem{grimaldi_annealing_2022}
A.~Grimaldi \emph{et~al.}, ``{Experimental Evaluation of Simulated Quantum Annealing with MTJ-Augmented p-bits},'' in \emph{International Electron Devices Meeting (IEDM)}, 2022, pp. 539--542.

\bibitem{aadit_tempering_2023}
N.~A. Aadit, M.~Mohseni, and K.~Y. Camsari, ``{Accelerating Adaptive Parallel Tempering with FPGA-based p-bits},'' in \emph{IEEE Symposium on VLSI Technology and Circuits (VLSI)}, 2023, pp. 1--2.

\bibitem{onizawa_hyperparameter_2023}
N.~Onizawa, K.~Kuroki, D.~Shin, and T.~Hanyu, ``{Local Energy Distribution Based Hyperparameter Determination for Stochastic Simulated Annealing},'' \emph{IEEE Open Journal of Signal Processing}, vol.~4, pp. 452--461, 2023.

\bibitem{rabaey_chandrakasan}
J.~M. {Rabaey}, A.~P. {Chandrakasan}, and B.~{Nikolic}, \emph{{Digital Integrated Circuits: A Design Perspective}}.\hskip 1em plus 0.5em minus 0.4em\relax Prentice-Hall, 2002.

\bibitem{weste_harris}
N.~H.~E. {Weste} and D.~M. {Harris}, \emph{{CMOS VLSI Design: A Circuits and Systems Perspective}}.\hskip 1em plus 0.5em minus 0.4em\relax Addison-Wesley, 2011.

\bibitem{ward_lfsr_2007}
R.~Ward and T.~Molteno, ``{Table of Linear Feedback Shift Registers},'' Department of Physics, University of Otago, Tech. Rep., Oct. 2007.

\bibitem{huang_tcas2_2023}
Z.~Huang, D.~Jiang, X.~Wang, and E.~Yao, ``{An Ising Model-Based Annealing Processor With 1024 Fully Connected Spins for Combinatorial Optimization Problems},'' \emph{IEEE Transactions on Circuits and Systems II: Express Briefs}, vol.~70, no.~8, pp. 3074--3078, 2023.

\bibitem{zhang_fpga_2022}
Y.~Zhang \emph{et~al.}, ``{Ultra High-Speed Polynomial Multiplications for Lattice-Based Cryptography on FPGAs},'' \emph{IEEE Transactions on Emerging Topics in Computing}, vol.~10, no.~4, pp. 1993--2005, 2022.

\bibitem{inagaki_cim_2016}
T.~Inagaki \emph{et~al.}, ``{A Coherent Ising Machine for 2000-Node Optimization Problems},'' \emph{Science}, vol. 354, no. 6312, pp. 603--606, 2016.

\bibitem{huang_iscas_2023}
Z.~Huang \emph{et~al.}, ``{An Annealing Processor based on 1k-Spin Fully-Connected Ising Model for Combinatorial Optimization Problems},'' in \emph{IEEE International Symposium on Circuits and Systems (ISCAS)}, 2023, pp. 1--5.

\bibitem{waidyasooriya_temporal_2022}
H.~M. Waidyasooriya and M.~Hariyama, ``{Temporal and Spatial Parallel Processing of Simulated Quantum Annealing on a Multicore CPU},'' \emph{The Journal of Supercomputing}, pp. 1--18, 2022.

\bibitem{yoshimura_architecture_2020}
C.~Yoshimura \emph{et~al.}, ``{CMOS Annealing Machine: A Domain-Specific Architecture for Combinatorial Optimization Problem},'' in \emph{Asia and South Pacific Design Automation Conference (ASP-DAC)}, 2020, pp. 673--678.

\bibitem{kulkarni_isingcim_2022}
S.~Xie \emph{et~al.}, ``{Ising-CIM: A Reconfigurable and Scalable Compute Within Memory Analog Ising Accelerator for Solving Combinatorial Optimization Problems},'' \emph{IEEE Journal of Solid-State Circuits}, vol.~57, no.~11, pp. 3453--3465, 2022.

\bibitem{kim_ctleising_2024}
J.~Bae \emph{et~al.}, ``{CTLE-Ising: A Continuous-Time Latch-Based Ising Machine Featuring One-Shot Fully Parallel Spin Updates and Equalization of Spin States},'' \emph{IEEE Journal of Solid-State Circuits}, vol.~59, no.~1, pp. 173--183, 2024.

\end{thebibliography}

\end{document}